\documentclass[prd,reprint,superscriptaddress,nobibnotes,nofootinbib]{revtex4-1}

\usepackage{aas_macros}
\usepackage{graphicx} 
\usepackage{amssymb}
\usepackage{amsmath}
\usepackage{subfigure}
\usepackage{enumitem}
\usepackage{color}

\begin{document}
\title{The diverse dark matter density at sub-kiloparsec scales in Milky Way satellites: implications for the nature of dark matter}

\author{Jes\'us Zavala}
\email[e-mail:]{jzavala@hi.is}
\affiliation{Center for Astrophysics and Cosmology, Science Institute, University of Iceland, Dunhagi 5, 107 Reykjavik, Iceland}
\author{Mark R. Lovell}
\affiliation{Center for Astrophysics and Cosmology, Science Institute, University of Iceland, Dunhagi 5, 107 Reykjavik, Iceland}
\affiliation{Institute for Computational Cosmology, Durham University, South Road, Durham DH1 3LE, UK}
\author{Mark Vogelsberger}
\thanks{Alfred P. Sloan Fellow}
\affiliation{Department of Physics, Kavli Institute for Astrophysics and Space Research, Massachusetts Institute of Technology, Cambridge, MA 02139, USA}
\author{Jan D. Burger}
\affiliation{Center for Astrophysics and Cosmology, Science Institute, University of Iceland, Dunhagi 5, 107 Reykjavik, Iceland}
\preprint{}
\date{\today}


\begin{abstract}
Milky Way (MW) satellites reside within dark matter (DM) subhalos with a broad distribution of circular velocity profiles. This diversity is enhanced with the inclusion of ultra-faint satellites, which seemingly have very high DM densities, albeit with large systematic uncertainties. We argue that if confirmed, this large diversity in the MW satellite population poses a serious test for the structure formation theory with possible implications for the DM nature. 
For the Cold Dark Matter model, the diversity might be a signature of the combined effects of subhalo tidal disruption by the MW disk and strong supernova feedback. For models with a dwarf-scale cutoff in the power spectrum,
the diversity is a consequence of the lower abundance of dwarf-scale halos. This diversity is most challenging for Self-Interacting Dark Matter (SIDM) models with cross sections $\sigma/m_\chi\gtrsim1~$cm$^2$g$^{-1}$ 
where subhalos have too low densities to explain the ultra-faint galaxies. 
We propose a novel solution to explain the diversity of MW satellites based on the gravothermal collapse of SIDM haloes. This solution requires a velocity-dependent cross section that predicts a bimodal distribution of cuspy dense (collapsed) subhaloes consistent
with the ultra-faint satellites, and cored lower density subhaloes consistent with the brighter satellites.
\end{abstract}


\maketitle

\section{Introduction}

The cold and collisionless nature of dark matter (DM) is a central hypothesis of the standard CDM model of structure formation, which is largely consistent with the large scale structure of the Universe 
\cite{Springel2005}, and is the cornerstone of current state-of-the-art simulations that model the complexity of galaxy formation and evolution \cite{Dubois2014,Vogelsberger2014,Schaye2015}. However, the validation of these hypotheses remains elusive since the predictions that make CDM a distinct model occur at (sub)galactic scales, where gas/stellar ({\it baryonic}) physics and possible new DM physics are entangled. It is in fact at these scales that CDM has been challenged over recent decades by its apparent inconsistency with observations, particularly by seemingly overpredicting the abundance and inner DM content of dwarf galaxies  \cite{Moore1994,Klypin1999,Moore1999,Zavala2009,Walker2011,Boylan-Kolchin2011,Klypin2014,Papastergis2015,Oman2015}. Whether this indicates the need for new DM physics or the lack of an accurate account of baryonic physics remains controversial. 
Supernova feedback and gas heating during the reionization era suppress the formation of dwarf galaxies and reduce their inner DM densities (e.g. \cite{Pontzen2012,Dutton2016,Sawala2016}), but there is no firm evidence of the high efficiency required from these processes to be viable solutions to all CDM challenges (see \cite{Bullock2017} for a review).

Among the range of allowed DM physics that can impact the physics of galaxies, there are two mechanisms that encompass a large set of possible DM particle models: (i) a dwarf-scale cutoff in the linear matter power spectrum, either caused by free streaming as in Warm Dark Matter (WDM) \cite{Colin2000,Bode2001}, or by interactions between DM and relativistic particles in the early Universe \cite{Boehm2002,Buckley2014}; (ii) a reduction of the central density of halos
if DM is self-interacting (SIDM, \cite{Spergel2000}). Although these two mechanisms naturally alleviate the CDM challenges (e.g. \cite{Lovell2012,Vogelsberger2012,Zavala2013,Vogelsberger2016,Schneider2017,Kamada2017}), verifying/falsifying them remains a challenge due to their interplay/degeneracies with baryonic physics (although see \cite{Burger2019}).  

In this work we revisit  the challenge of matching the abundance and kinematic properties of the Milky Way (MW) satellites by looking at the so-called too-big-to-fail (TBTF) problem, which states that the most massive subhaloes predicted by CDM $N$-body simulations are too centrally dense to host the brightest MW satellites  \cite{Boylan-Kolchin2011,Boylan-Kolchin2012}. We take a different perspective of the TBTF challenge in light of recent observations of ultra-faint galaxies, which indicate a strikingly diverse distribution of the internal kinematics of this dispersion-supported satellite population \cite{Fattahi2018,Errani2018}. Such diversity is akin to the diversity of rotation curves reported in higher mass, rotationally-supported dwarf galaxies \cite{Oman2015}. 
Our goal is  to show how this diversity in MW satellites poses a serious test for structure formation models and focus particularly on its implications for the DM nature.

The paper is organized as follows. In Section \ref{sec_simulations} we describe the DM models we study and their corresponding cosmological simulations. In Section \ref{sec_results} we present our results on the distribution of circular velocity profiles in the simulated subhalo population and its comparison with observations from the MW satellites. In Section \ref{sec_discussion} we discuss a number of factors that impact our results as well as discuss the crucial role that the gravothermal collapse of SIDM halos could have for the SIDM model. Finally, our Conclusions are given in Section \ref{sec_conclusions}.

\section{DM models and Simulations}\label{sec_simulations}

\begin{table*}
\begin{center}
\begin{tabular}{ccccccccccc}
\hline
Name     & Cosmology  &  $\sigma_T/m_\chi$ & $m_{\rm WDM}$ & $M_{\rm 200}$           & $m_{\rm DM}$                   & $\epsilon$  & Reference &  Aquarius IC\\
            & 		& 	[cm$^2$g$^{-1}$]	&	[keV] &  $[10^{12}~{\rm M_\odot}]$          & $[{\rm M_\odot}]$     & $[{\rm pc}]$  &    &    \cite{Springel2008}  \\
\hline     
\hline 
CDM      & WMAP-7 & $-$ & $-$   & 1.94                                    & 1.55$\times10^4$                          & 68.1          & \cite{Lovell2014}      &    Aq-A-2 \\
SIDM     & WMAP-1 & $1.0$ & $-$ & 1.80                                    & 4.9$\times10^4$                          & 120.5          & \cite{Zavala2013}     &    Aq-A-3  \\
WDM     & WMAP-7 & $-$ & $2.32$ & 1.87                                    &  1.55$\times10^4$                          & 68.1         & \cite{Lovell2014}      &    Aq-A-2  \\
ETHOS     & Planck$-$2015 & 0.3 (v$_{\rm rel}=10$~km/s)& 3.4 & 1.64    & 2.76$\times10^4$   & 72.4         & \cite{Vogelsberger2016}    &    $-$   \\
vdSIDM     & WMAP-1 & 125 (v$_{\rm rel}=10$~km/s) & $-$ &           1.84                          & 4.9$\times10^4$                          & 120.5          &  this work    &   Aq-A-3    \\
\hline
\end{tabular}
\end{center}
\caption{{\it Main MW-size zoom-in simulation properties.} Columns: (1) label; (2) the dataset to which the cosmological parameters in each simulation are consistent with: WMAP 1st year \cite{Spergel2003}, WMAP 7th year \cite{Komatsu2011} or Planck 2015 \cite{Planck2016}; (3) the transfer cross section per unit mass for the SIDM, ETHOS and vdSIDM cases; the latter two are velocity-dependent, and only the value at 10 km/s is shown as reference; (4) the thermal relic mass for the WDM  and ETHOS cases; the latter is not a WDM model, but it has a cutoff to the linear power spectrum with a scale similar to a thermal relic WDM model with the mass shown on the table (this model is labeled ETHOS-4 in \cite{Vogelsberger2016}; see Fig. 1 and Table 1 therein, and Fig. 1 in \cite{Lovell2019}); (5) the mass within a radius enclosing a mean density of 200 times the critical density; (6) the simulation particle mass; (7) the Plummer-equivalent softening length; (8) the original reference describing each simulation; (9) the reference to the initial conditions for the zoom-in  simulations. With the exception of the ETHOS simulation, all other cases correspond to the simulation volume labeled Aq-A in the Aquarius simulation suite \cite{Springel2008}.}
\label{table:sims}
\end{table*}

\begin{figure*}
 \subfigure{\includegraphics[height=8.0cm,width=8.5cm,trim=1.25cm 0.5cm 0.0cm 0.0cm, clip=true]{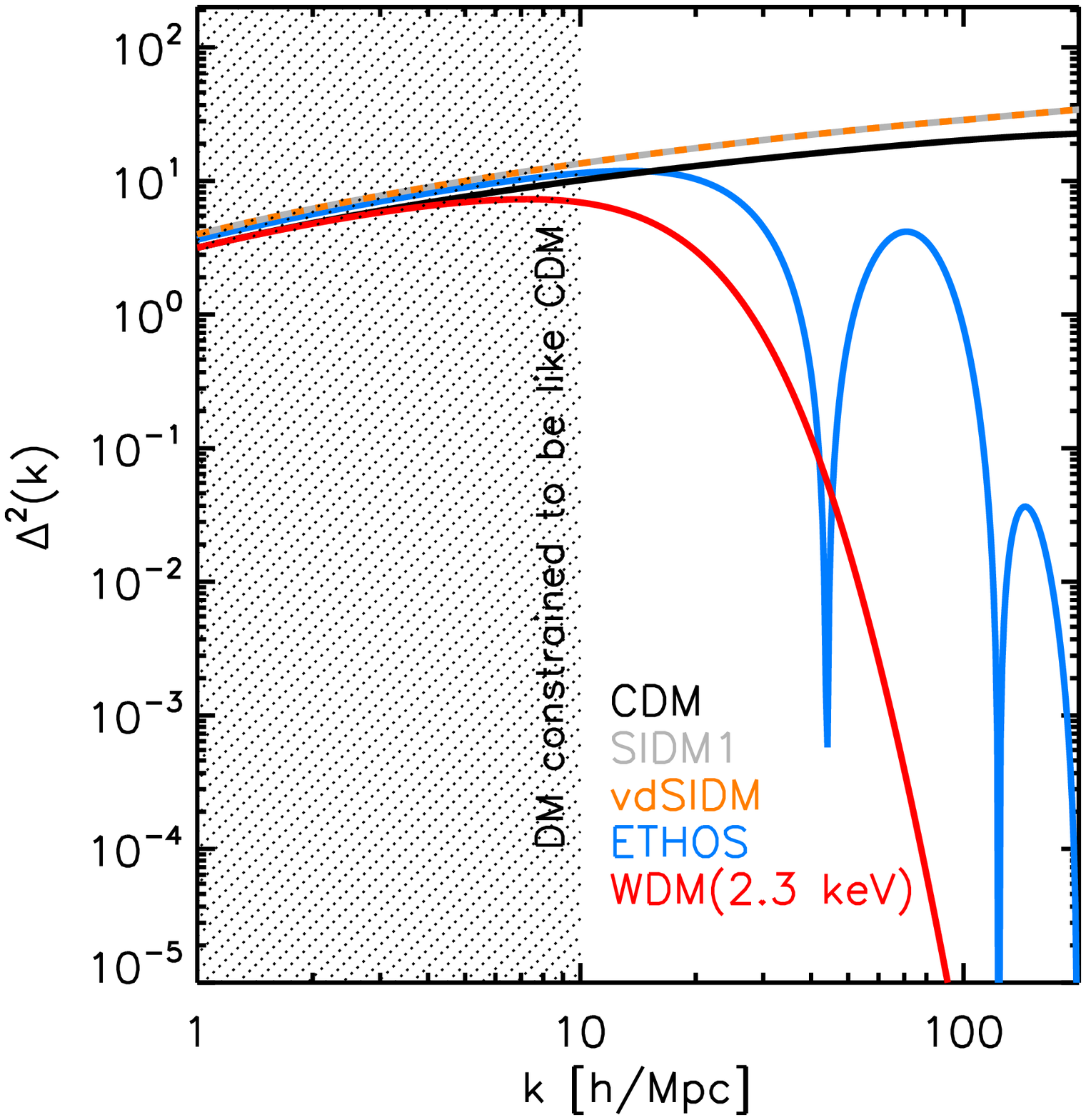}} 
  \subfigure{\includegraphics[height=8.0cm,width=8.75cm,trim=0.25cm 0.5cm 0.75cm 0.0cm, clip=true]{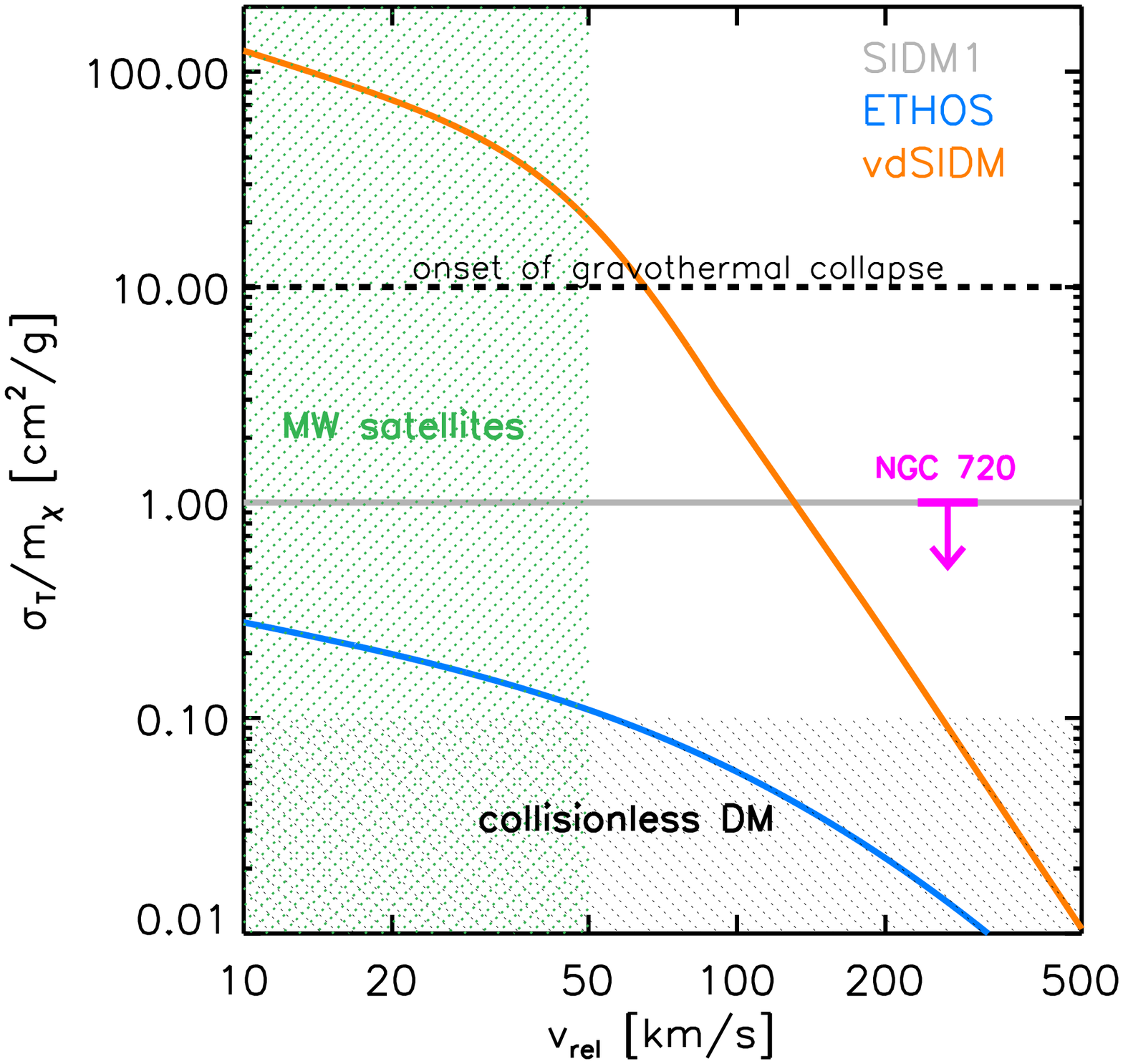}}
\caption{{\it Left:} Dimensionless linear DM power spectra ($\Delta^2(k)=k^3P_{\rm linear}(k)/2\pi^2$) for the simulations
used in this work. The CDM (black), SIDM (gray), and vdSIDM (orange) cases have a CDM-like power spectrum 
while the WDM (red) and ETHOS (blue) cases have galactic-scale cutoffs produced by free streaming and DM$-$dark radiation interactions in the early Universe, respectively. In the black hashed area, DM is constrained to behave like CDM (e.g. Ly-$\alpha$ constraints on WDM \cite{Viel2013}). Notice that not all simulations have the same cosmological parameters, and thus there is a mismatch at large scales. {\it Right:} The transfer cross section as a function of relative velocity for the simulations with self-interactions.
The black hashed area with $\sigma_T/m_\chi<0.1$~cm$^2$g$^{-1}$ marks the region where DM is effectively collisionless. The green hashed area on the left is the relevant region for MW satellites. Above the dashed line, self-interactions are frequent enough for the onset of gravothermal collapse within a Hubble time in SIDM halos. The magenta arrow is the limit to the cross section from shape measurements of the elliptical galaxy NGC720 \cite{Peter2013}.}
\label{fig:models}
\end{figure*}

We analyse a sample of DM-only cosmological simulations of MW-size halos, four of which have been used in the past to study subhalo abundance and their inner structure (see Table~\ref{table:sims}). 
Fig.~\ref{fig:models} shows the linear power spectra (left) and self-interaction cross sections (right) for the different DM models used in these simulations. The motivation for using this particular set of simulations is to have a sample with sufficient and roughly similar resolution of $\mathcal{O}$(100~pc) to probe the inner region of MW satellites in different DM models: CDM \cite{Lovell2014}, WDM ($m_{\rm WDM}=2.3$~keV; \cite{Lovell2014}), SIDM ($\sigma_T/m_{\chi}=1$~cm$^2$g$^{-1}$; \cite{Zavala2013}), and a model within the ETHOS framework of structure formation \cite{Cyr2016}. The latter is a benchmark case fine-tuned in \cite{Vogelsberger2016} to alleviate the TBTF and abundance problems. It contains both self-interactions ($\sigma_T/m_\chi\sim0.3$~cm$^2$g$^{-1}$ at the characteristic velocities of MW satellites) and a primordial cutoff in the power spectrum (nearly equivalent to a $\sim3.4$ keV WDM model; \cite{Lovell2019}) due to DM$-$dark radiation interactions in the early Universe. 

We notice that the 2.3 keV thermal WDM model we use is disfavored at $>3\sigma$ C.L. from observations of the Lyman$-\alpha$ forest flux power spectra \cite{Viel2013,Irsic2017}. 
Although it has been pointed out that uncertain factors in the high redshift Universe, most notably the thermal history of the intergalactic medium, could greatly relax the constraint on the WDM particle mass (e.g \cite{Garzilli2017}), very recent inferences on this thermal history \cite{Walther2018} reduce this possibility and seemingly validate the constraints in \cite{Viel2013,Irsic2017}. 
In the ETHOS model we consider however, the power spectrum cutoff occurs at smaller scales (roughly analogue to a 3.4 keV thermal WDM model; \cite{Lovell2019}), and thus it is in considerable less tension with Ly-$\alpha$ forest data (see \cite{Bose2018}). 

On the other hand, most limits on the self-interacting transfer cross section are in place for systems with characteristic velocities larger than those in the MW satellites and are of the order of $\sigma_T/m_{\chi}=1$~cm$^2$g$^{-1}$ (for a review of the constraints see Table 1 of \cite{Tulin2018}). A recent study reports a constraint of $\sigma_T/m_{\chi}<0.57$~cm$^2$g$^{-1}$ ($99\%$ C.L.) at precisely the velocity scales of MW satellites through measurements of the inner DM density in the Draco satellite (\cite{Read2018}; see also \cite{Valli2018}). Although this is potentially a relevant constraint on SIDM, Draco has a stellar mass of $\sim3\times10^5$~M$_\odot$ and it cannot be ruled out that baryonic physics could impact
the inner profile of low-mass SIDM halos. In fact, for galaxies with $\sim10^6$~M$_\odot$ in stellar mass, it has been shown that the profiles of SIDM halos are cuspier than their DM-only counterparts (see Fig. 5 of \cite{Robles2017}).

To this simulation suite we add a new one (vdSIDM) with the same initial conditions as the SIDM simulation but with a strong velocity dependent cross section (orange line in the right panel of Fig.~\ref{fig:models}). In the particular model we use, 
DM self-scattering is mediated by a massive force carrier ($m_\phi$)  through an attractive Yukawa potential with coupling strength $\alpha_c$. In this case, the transfer cross section in the classical regime can be approximated by a fitting function used in plasma physics (see e.g. \cite{Feng2010,Loeb2011,Vogelsberger2012}):
\begin{equation}
\frac{\sigma_T}{\sigma_T^{\rm max}} \approx 
     \begin{cases}
       \frac{4\pi}{22.7}~\beta^2~{\rm ln}\left(1+\beta^{-1}\right),                    &\beta<0.1\\ \\
       \frac{8\pi}{22.7}~\beta^2~\left(1+1.5\beta^{1.65}\right)^{-1},                   &0.1<\beta<10^3\\ \\
       \frac{\pi}{22.7}~\left({\rm ln}\beta+1-\frac{1}{2}{\rm ln}^{-1}\beta\right)^2,  &\beta>10^3, 
     \end{cases}
\label{eq:cross}
\end{equation}
where $\beta=\pi v_{\rm max}^2/v^2=2\,\alpha_c\,m_{\phi}/(m_{\chi}v_{\rm rel}^2)$ and $\sigma_T^{\rm max}=22.7/m_{\phi}^2$, and $v_{\rm rel}$ is the
relative velocity of the DM particles. Here $v_{\rm max}$ is the velocity at which $(\sigma_T v_{\rm rel})$ peaks at a transfer cross
section equal to $\sigma_T^{\rm max}$. We choose the particle physics parameters so that $v_{\rm max}=25$~km/s and $\sigma_T^{\rm max}=60$~cm$^2$g$^{-1}$. With this choice of parameters, 
self-interactions are frequent enough in the center of dwarf-scale SIDM (sub)halos to trigger the gravothermal catastrophe phase, which is a well-known mechanism in globular clusters \cite{LyndenBell1968}, and that in the SIDM case results in the collapse of the core into a central cusp \cite{Balberg2002,Colin2002,Koda2011,Pollack2015,Nishikawa2019}. 

\section{Results}\label{sec_results}

\begin{figure*}
    \subfigure{\includegraphics[height=9.25cm,width=9.25cm,trim=1.75cm 2.5cm 1.15cm 0.0cm, clip=true]{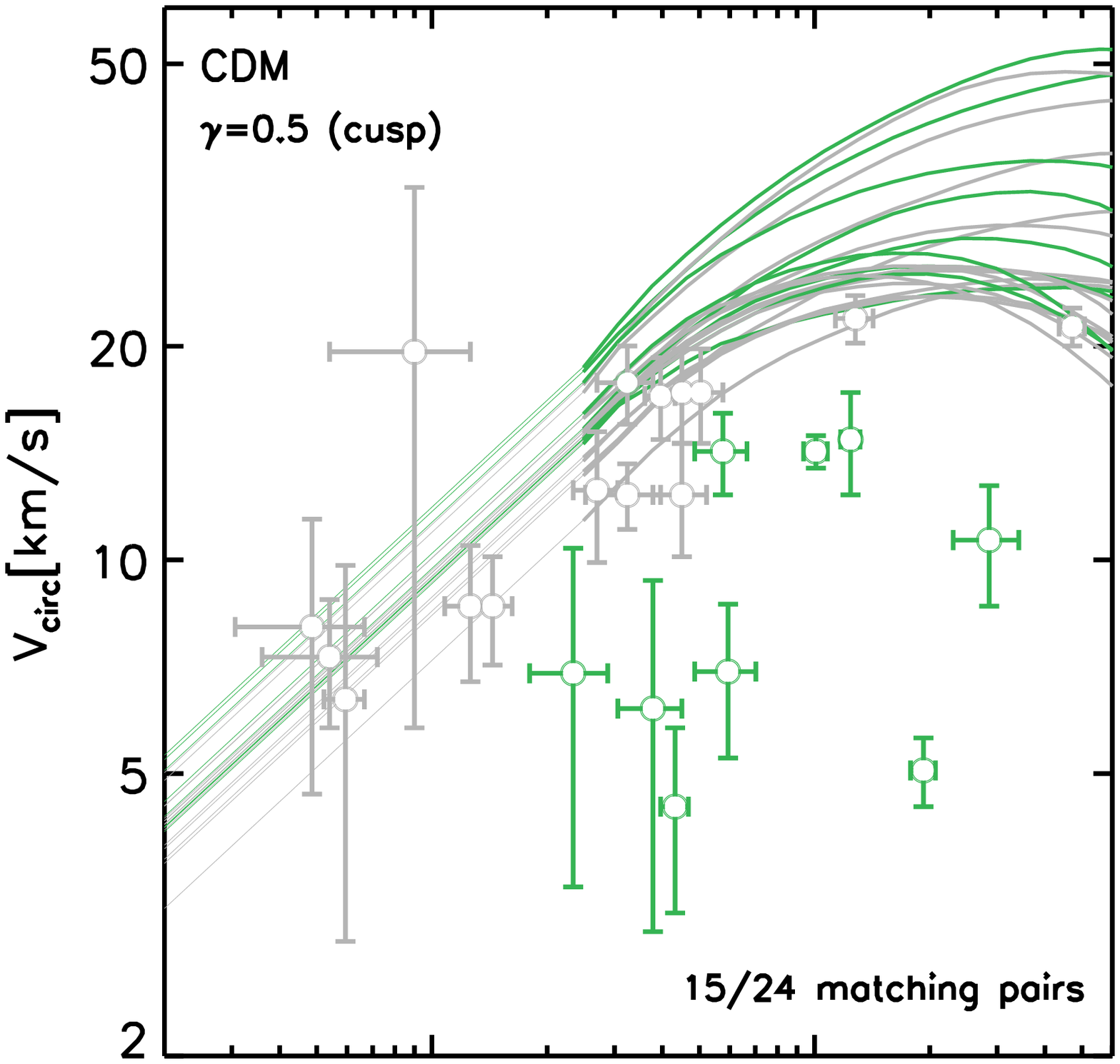}} 
    \subfigure{\includegraphics[height=9.25cm,width=8.5cm,trim=4.15cm 2.5cm 0.75cm 0.0cm, clip=true]{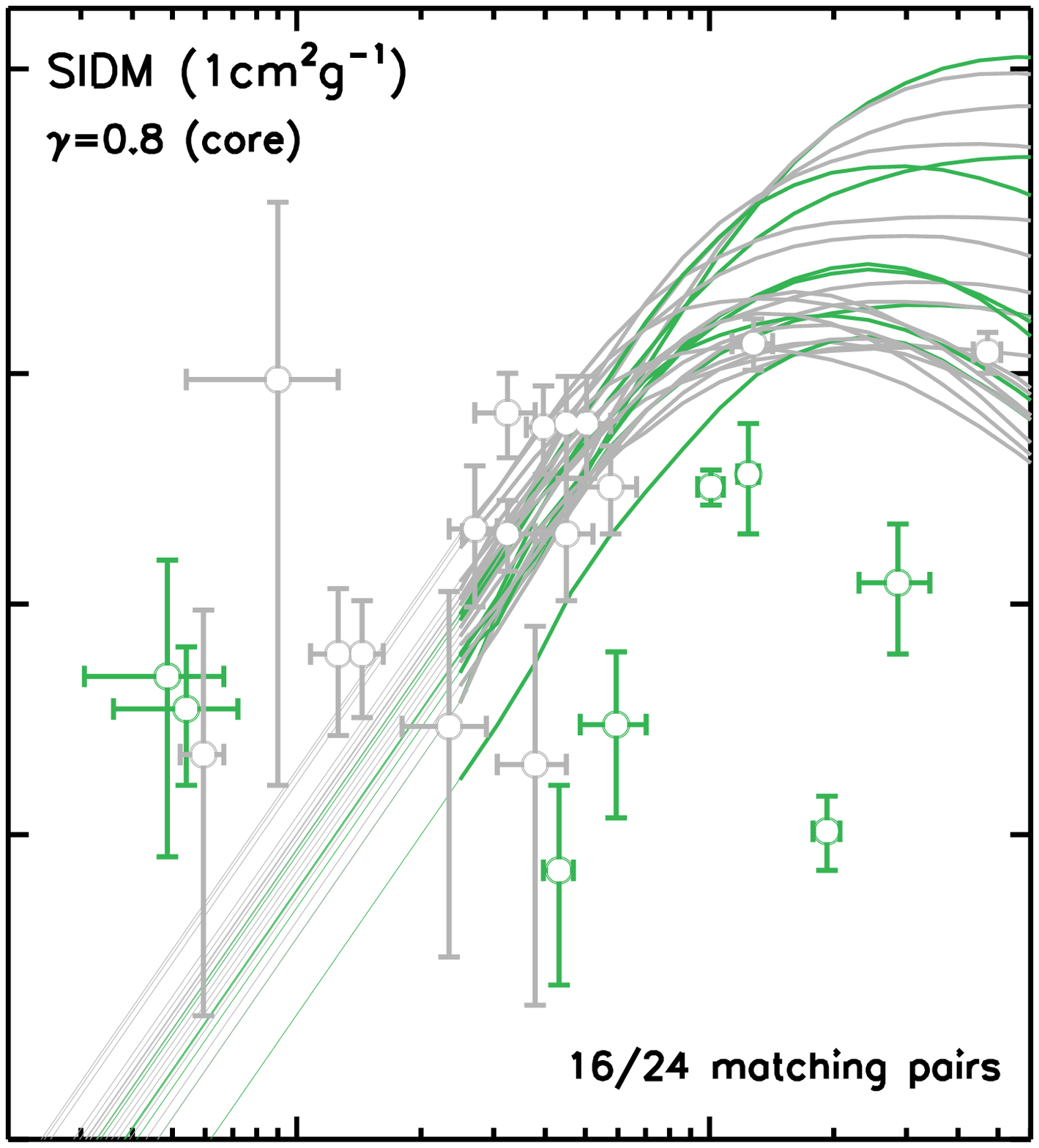}}
    \hfill 
    \subfigure{\includegraphics[height=9.25cm,width=9.25cm,trim=1.75cm 0.7cm 1.15cm 1.25cm, clip=true]{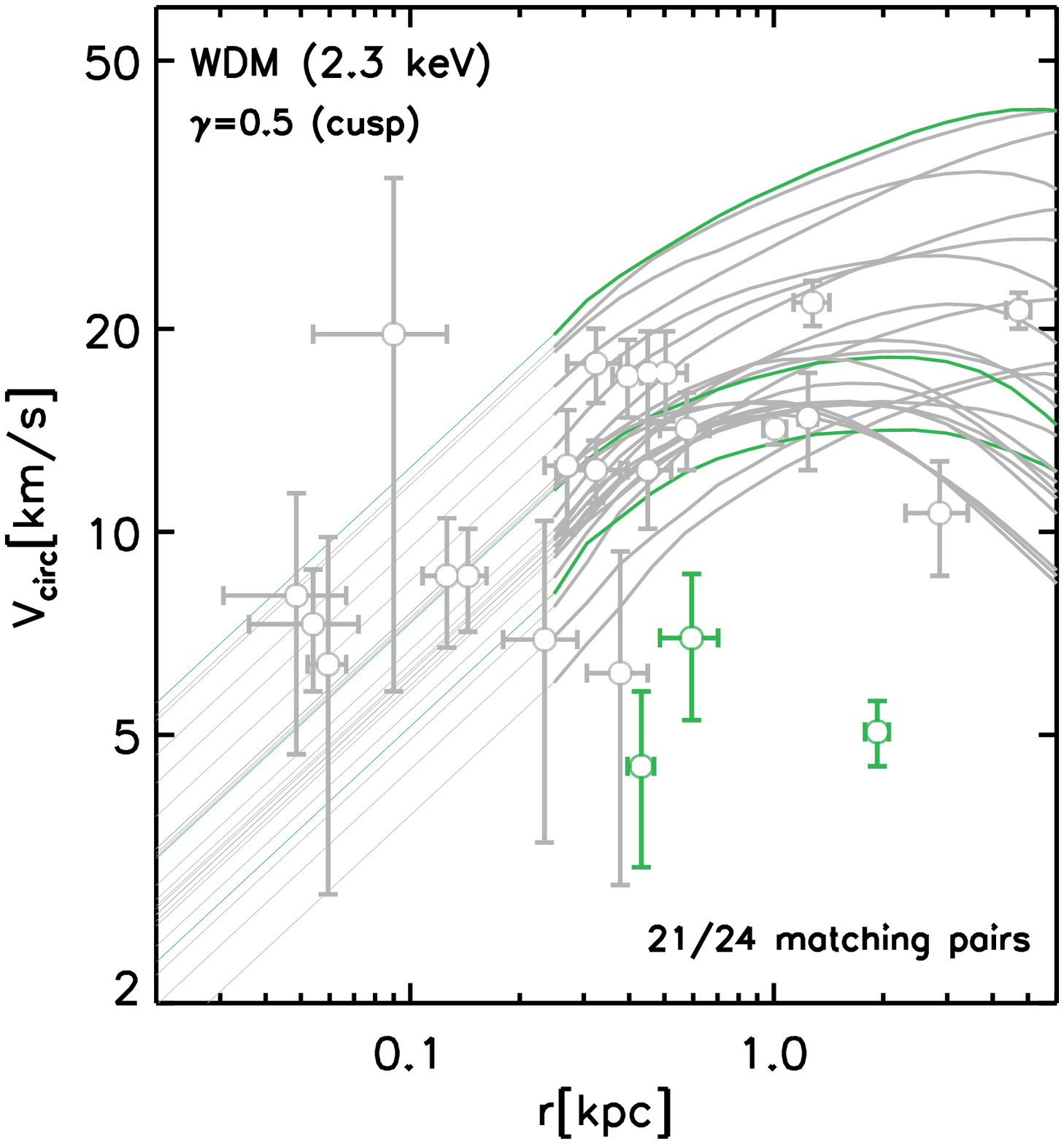}}
    \subfigure{\includegraphics[height=9.25cm,width=8.5cm,trim=4.15cm 0.7cm 0.75cm 1.25cm, clip=true]{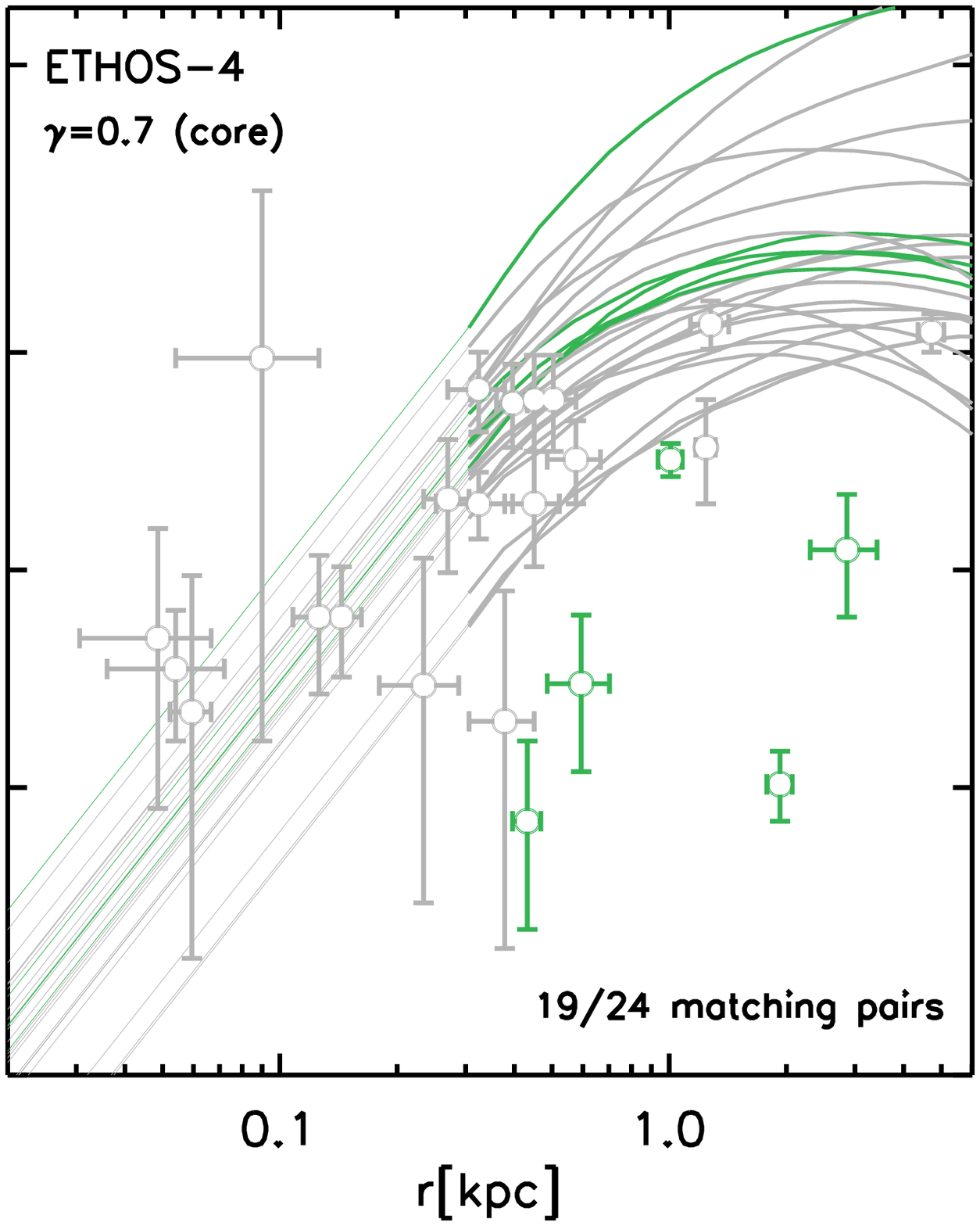}}
\caption{The circular velocity profiles for the 24 subhalos with the largest values of $V_{\rm max}(z=0)$ within $300$~kpc of the center of the MW-size halo, after excluding Magellanic Cloud analogs. 
We show four different cosmologies (see Fig.~\ref{fig:models}): CDM, SIDM with $\sigma_T/m_\chi=1$~cm$^2$g$^{-1}$, WDM with $m_\chi=2.3$~keV, and a benchmark model within the ETHOS framework, which has self-interactions and a primordial power spectrum cutoff \cite{Cyr2016,Vogelsberger2016}. The solid lines show the profiles beyond the convergence radius. 
Below this radius (thin lines), most subhalos have reached the expected asymptotic values: $V_{\rm circ}\propto r^\gamma$ for cuspy (left panels) and cored profiles (right panels). 
Open symbols with error bars show $V_{\rm circ}$ values at the half light radius for 24 MW satellites as given in \cite{Errani2018,Torrealba2018}. 
Lines and symbols in gray are examples of consistent matches of simulated subhalos and data points (the largest possible number of matching pairs is shown in the lower right). The mismatches are shown in green.}
\label{fig:vcirc}
\end{figure*}

In Fig.~\ref{fig:vcirc} we compare the distribution of the circular velocity profiles, $V_{\rm circ}(r)$, of the massive subhalo population in the different DM models with the observed distribution of circular velocities at the half light radius $V_{1/2}\equiv V_{\rm circ}(r_{1/2})$ for the MW satellite population. 
We have taken the observed values of $V_{1/2}$ and $r_{1/2}$ from \cite{Errani2018}, where a new mass estimator to infer the dark matter content in pressure-supported systems was employed that the authors claim provides unbiased mass estimates with a $\sim10\%$ accuracy. Using this estimator we have:
\begin{equation}
V_{1/2}^2=V^2_{\rm circ}(r_{1/2}=1.8R_{1/2})=3.5\langle\sigma^2_{\rm los}\rangle
\end{equation}
where $R_{1/2}$ and $\langle\sigma^2_{\rm los}\rangle$ are the observed projected half-light radius and luminosity-averaged squared line-of-sight velocity dispersion of the stars (given in Table 2 of \cite{Errani2018}). We also include the newly discovered Antlia 2 satellite \citep{Torrealba2018} with $(r_{1/2},V_{1/2})$ values computed with the same estimator as in \cite{Errani2018}. 

We only consider satellites (subhalos) within 
300 kpc from the center of the observed (simulated) MW system. This leaves us with 24 confirmed MW satellites (after excluding Leo T), which we compare against the 24 subhalos in each simulation with the largest maximum circular velocity $V_{\rm max}(z=0)$, excluding the two subhalos with the highest ranking, which are deemed to be Magellanic Cloud analogs. The simulation data in Fig.~\ref{fig:vcirc} extend down to a radius where convergence is reasonable (solid lines): $\sim3.5\epsilon$ for the CDM, WDM and ETHOS models, and $\sim2\epsilon$ for the SIDM case. The latter is smaller than in the other cases because SIDM halos are much better converged in the central regions due to the thermalization of the core \cite[e.g.][]{Vogelsberger2012,Elbert2015}. 

The gravitational softening lengths of the simulations are of $\mathcal{O}$(100 pc), which is around the scale of the half-light radii of the ultra-faint galaxies. In order to extrapolate the simulation results to this unresolved regime we proceed as follows.
For the CDM and WDM cases, we note that the simulations are sufficiently resolved down to $\sim200$~pc to approach the expected asymptotic value of the NFW profile: $V_{\rm circ}\propto r^{1/2}$ \cite{Lovell2014}. In fact, we have verified that power law fits to the last 4 resolved radial bins in the CDM and WDM cases result in a median slope that is very close to $0.5$. Because of this, we extrapolate the $V_{\rm circ}$ profiles for these simulations (left panels of Fig.~\ref{fig:vcirc}) with power laws of fixed NFW slope and a normalization given by the last resolved radial bin. 
For the SIDM and ETHOS simulations (right panels of Fig.~\ref{fig:vcirc}) we proceed in a similar way, fixing the slopes of the power laws to the median slope of the last 4 resolved radial bins of the $V_{\rm circ}$ profiles. In this case, we find that at $r\gtrsim200$~pc the profiles have not yet reached the asymptotic value for a profile with a flat central core: $V_{\rm circ}\propto r$. Instead, the SIDM case has $V_{\rm circ}\propto r^{0.8}$ while the ETHOS case has $V_{\rm circ}\propto r^{0.7}$. The reason for this is that the thermalization of the core is neither perfect nor complete, particularly for the ETHOS case where the cross section for the typical velocities of the considered subhalo population at these scales ($\sim10$~km/s) is $\sim0.3$~cm$^2$g$^{-1}$ (see right panel of Fig.~\ref{fig:models}), resulting in an isothermal region that is much smaller than the maximum it can attain. Finally, for the vdSIDM simulation (Fig.~\ref{fig:vcirc_vdsidm} in Section \ref{sec_gravothermal}), we separate the subhalo population into two distinct subpopulations according to the behavior of their circular velocity profiles in the last resolved radial range: $V_{\rm circ}\propto r^\gamma$. If $\gamma\geq0.65$, then subhalos are cored-like, otherwise they are cusp-like. We then proceed as in previous cases and assign to each subclass an asymptotic behavior given by the median value of $\gamma$ computed for each subclass. Notice that although the specific value of $\gamma$ used to divide the population is somewhat arbitrary, it serves the purpose of characterizing the bimodal distribution of $V_{\rm circ}$ profiles that is apparent in the vdSIDM case.

Although the spread and normalization of the distributions give an idea of how discrepant/similar the simulations are to the data, we can proceed further and establish a consistency between the simulations and the data by
defining matching pairs, a circular velocity profile and a ($r_{1/2}$,$V_{1/2}$) observational point, and
finding the largest set of subhalos that is consistent 
with the largest number of data points in the following way. 
In order to be conservative in our assessment of the consistency between the simulations and the data, our goal is to find the maximum number of possible matching simulated subhalo$-$observed satellite pairs. To accomplish this we use the following procedure. For each data point, we find all subhalos with a circular velocity profile that lie within a square defined by the error bars of that data point. Once this list of possible matches is built, we then randomly choose matching pairs by first selecting at random a data point and then at random a matching subhalo. Both of these random choices are sampled from uniform distributions (without replacement) and the process is repeated 1000 times, which is sufficient to ensure that the maximum possible number of matches is achieved. A random realisation that achieves the maximum number of matches is chosen as the example shown in Figs.~\ref{fig:vcirc} and \ref{fig:vcirc_vdsidm} for each DM model.

The matching pairs (mismatches) are shown in Fig.~\ref{fig:vcirc} in gray (green). 
The upper panels of Fig.~\ref{fig:vcirc} show the cases that are more discrepant with the data: CDM (left) and SIDM (right). CDM has the well-known TBTF problem, having too many dense subhalos to explain the satellite distribution.
On the other hand, SIDM with 1~cm$^2$g$^{-1}$ predicts subhalos with too low densities to match the ultra-faint galaxies. 
In both the CDM and the SIDM models, the spread of the profile distribution is a problem: the subhalo population is too narrow to account for the large spread in the data. This problem is alleviated without appealing to baryonic physics if the subhalo abundance is suppressed due to a primordial power spectrum cutoff, regardless of whether subhalos are cored or cuspy. This is shown in the bottom panels of Fig.~\ref{fig:vcirc}, where both the WDM and ETHOS models show a more reasonable match to observations.

\section{Discussion}\label{sec_discussion}

In the following we discuss a number of limitations/circumstances of the 
simulations we use, which impact the results shown in Fig.~\ref{fig:vcirc}: choice of cosmological parameters, MW halo mass and halo-to-halo environmental scatter, choice of subhalo ranking, as well as the most relevant (baryonic) physical processes not present in our simulations: adiabatic contraction, supernova feedback and tidal disruption by the MW disk.

\begin{itemize}[leftmargin=0.5cm]

\item[a)]{\it Cosmological parameters.} The different parameters used in the simulations have
an impact on the inner densities of (sub)halos. In particular, a larger value of $\sigma_8$, the rms amplitude of linear mass fluctuations in 8 Mpc/h spheres at redshift zero, results in satellites forming earlier and therefore being denser, which makes the too-big-to-fail problem more severe \cite{Polisensky2014}. Qualitatively, this is only relevant for the SIDM and vdSIDM simulations, which have an overly high value of $\sigma_8$. Because of this, we would expect
SIDM to be even more discrepant with the ultra-faint galaxies if the simulation were repeated with a lower more consistent $\sigma_8$ value.

\item[b)]{\it MW halo mass and halo-to-halo scatter.} All of the simulations in our sample have a MW halo at the extreme high mass end of current estimates \cite{Wang2015,Patel2018}. 
On the other hand, since we are looking at a single realization of a MW-size halo, we are not considering the possible 
variations in the subhalo populations at a fixed halo mass. We remark however, that except for the ETHOS case, the particular halo we are considering is part of the haloes belonging to the Aquarius simulation suite 
(Aq-A; \citep{Springel2008}), which have been found to be fairly representative of the MW-like halo population in the CDM cosmology \citep{BK2010}, particularly with a subhalo abundance that is similar or larger than the cosmological median \citep{Cautun2014}. Taking this into consideration and acknowledging that a less massive halo would result in an overall less dense subhalo population, we expect Fig.~\ref{fig:vcirc} to be shifted downwards for all models. The consequences would be more striking for the SIDM case, as it would amplify its discrepancy with the data. For our conclusions to change significantly, the actual MW halo would likely need to deviate significantly from the median expectations with a larger abundance of dense subhalos than expected. 

\item[c)]{\it Subhalo ranking.} For simplicity, we have chosen $V_{\rm max}(z=0)$ to rank subhalos, but given the environmental effects (e.g. tidal and ram pressure stripping) that affect a subhalo once it is accreted into a larger host, $V_{\rm max}(z_{\rm infall})$, with $z_{\rm infall}$ being the infall time into the MW, would be a more appropriate choice theoretically due to its stronger correlation with satellite luminosity (see e.g. Fig. 6 of \cite{Guo2015}).
Unfortunately, we do not have the subhalo assembly histories for all the simulations to compute $V_{\rm max}(z_{\rm infall})$, but we note that the expectation in this case is that a few of the subhalos towards the lower end of the distribution of $V_{\rm circ}$ profiles would be replaced by subhalos that have suffered from more severe tidal stripping and that, at $z=0$, have $V_{\rm circ}$ profiles that are lower than those shown in Fig.~2. The distribution towards the massive (higher) end would remain essentially unchanged since, in general, subhalos that are massive at $z=0$ were also massive in the past (e.g. see Fig. 2 of \cite{BK2012}). We have explicitly verified this expectation for those simulations where we have the values of $V_{\rm max}(z_{\rm infall})$ (CDM and WDM). Thus, the net result is that ranking subhalos at infall would result in a slightly more diverse distribution in Fig.~\ref{fig:vcirc}, resulting in an overall better match to observations for all models. It would not however, alleviate the discrepancy of the SIDM model with the ultra-faint galaxies.

\item[d)]{\it Adiabatic contraction within subhalos.} The assembly of the galaxy leads to an adiabatic contraction of the DM halo, which makes its density profile steeper towards the center \cite{Barnes1984,Blu1986}. The relevance of this effect depends on the total mass of the galaxy and its concentration relative to those of the host halo. This mechanism could enhance the diversity of $V_{\rm circ}$ profiles only if the subhalo hosts of the ultra-faint galaxies were to become denser as a result. However, looking at Fig.~2 these galaxies are most likely hosted by the most massive MW subhalos, which are
too massive ($\gtrsim10^9$~M$_\odot$ at $z=0$) to be affected by adiabatic contraction since the ultra-faint galaxies have stellar masses $\lesssim10^4$~M$_\odot$. The effect could be more relevant for the cored SIDM subhalos, and thus we explicitly simulated this effect with a higher resolution, controlled simulation of an isolated SIDM halo. This halo has similar properties to the most massive SIDM subhalos, with a ``galaxy'' that is dynamically modeled with an external Plummer potential, and has a total stellar mass and half light radius characteristic of the ultra-faint galaxies. We found no significant impact on $V_{\rm circ}(r)$ down to the resolved scales $\sim60$~pc.

\item [e)]{\it Supernova feedback and tidal disruption by the MW disk.} The (gravitational) transfer of energy to the DM particle orbits by supernovae lowers the density of DM halos with a strength that strongly depends on the stellar-to-halo mass ratio (e.g. \cite{diCintio2014}), being largely inefficient for systems such as the ultra-faint galaxies (if they indeed live within massive subhalos). On the other hand, the presence of a MW disk can lower the densities of MW subhalos and even destroy them if the pericenter of their orbits gets too close to the disk (e.g. \cite{Zolotov2012}). A recent cosmological hydrodynamical simulation of the MW halo and its local environment shows that these mechanisms (with tidal disruption likely being the most relevant one) naturally create a more diverse subhalo population (relative to the CDM predictions without baryonic physics), and alleviate the too-big-to-fail problem \cite{GK2018} (see also \cite{Sawala2016,GK2017,Fattahi2018}).

\end{itemize}

We argue that all of these considerations are unlikely to modify the discrepancy of the 1~cm$^2$g$^{-1}$ SIDM model with the ultra-faint galaxies. Baryonic physics (particularly tidal disruption by the MW disk) would however, 
naturally enhance the diversity of $V_{\rm circ}$ profiles in the MW satellites \cite{Lovell2017}. In the case of CDM this results in a galaxy population that is seemingly a good match to the distribution of the {\it classical} dwarf satellites in the MW (see Fig. 5 of \cite{GK2018}); the ultra-faint galaxies remain unresolved in full cosmological hydrodynamical simulations, so it remains unclear how well they are described in CDM. For non-CDM models, it remains to be studied in detail how tidal disruption by the MW disk combines with (i) a cutoff in the power spectrum at the scale of MW satellites (such as in the WDM and ETHOS models) and/or (ii) SIDM-induced $\mathcal{O}$(kpc) cores within MW satellites, to enhance the diversity in DM densities within sub-kiloparsec scales. 
If this combination overfixes the problem, it could put stringent constraints on non-CDM models. 

In regards to point ii) in the paragraph above, \cite{Robles2019} recently performed MW-size CDM and SIDM ($1$~cm$^2$g$^{-1}$) simulations, which model the tidal effect of the MW disk by adding an embedded time-dependent potential. This changes substantially the DM distribution of the {\it host MW halo} relative to the DM-only simulation without the effect of the disk (making it much more centrally dense than even the CDM counterpart). However, the sub-kpc distribution of DM densities of the most massive subhaloes is altered only in a 
minimal way, relative to the DM-only simulation (see Fig. 3 of \cite{Robles2019}). Overall the SIDM subhaloes are slightly more dense than in the case without the effects of the disk, but still clearly less dense than in the CDM case. Based on this result, our expectation for the SIDM case with $1$~cm$^2$g$^{-1}$ remains in regards to its discrepancy with the ultra-faint galaxies. It would nevertheless be interesting to analyze why the MW tidal effects would enhance the SIDM densities for that particular simulation, although this is perhaps due to gravothermal collapse, which is accelerated when tidal stripping occurs, as was proposed recently by \cite{Nishikawa2019} (see also  \cite{Kahlhoefer2019,Sameie2019}).

\subsection{Systematic uncertainties in ultra-faint galaxies} \label{sec_uncertainties}

We stress that it is the large central densities 
of the ultra-faint galaxies that make the diversity of the satellite population particularly challenging. If the uncertainties on the mass estimator used in \cite{Errani2018} for these galaxies have been underestimated, this would relax the discrepancy noted here for the $1$~cm$^2$g$^{-1}$ SIDM model.  
The four ultra-faints we show in Figs.~\ref{fig:vcirc} and \ref{fig:vcirc_vdsidm} could have relevant systematic uncertainties: (i) the stellar kinematics data from Segue I is based on $\sim70$ stars \cite{Simon2011}, but not all of them are unambiguously identified as belonging to the satellite, which could affect the measurement of $\langle\sigma_{\rm los}\rangle$ (see Fig. 5 of \cite{Bonnivard2016}); (ii) the data for Willman 1 is based on $\sim15$ stars \cite{Martin2007} and a more detailed study with a larger sample \cite{Willman2011} suggested that Willman 1 might not be in dynamical equilibrium, and also found possible interlopers in the sample in \cite{Martin2007}, which might have biased high the value of  $\langle\sigma_{\rm los}\rangle$; (iii) the cases of Segue II and Bo\"otes II are even more uncertain with kinematics based only on a handful of stars \cite{Belokurov2009,Koch2009}; a study by \cite{Kirby2013} with $\sim20$ members of Segue II was not able to measure the velocity dispersion, and instead set an upper limit of $\langle\sigma_{\rm los}\rangle<2.6$~km/s (95$\%$ confidence), which remains consistent but at the lower end of the error bars reported in \cite{Belokurov2009}, while a study by \cite{Ji2016} on Bo\"otes II indicates that the velocity dispersion reported in \cite{Koch2009} might be biased high due to the inclusion of a star that is likely part of a binary system.

\subsection{Gravothermal collapse in SIDM halos} \label{sec_gravothermal}

The addition of baryonic physics
points towards a subhalo population that should diversify and move systematically towards lower $V_{\rm circ}$ values relative to the one shown in Fig.~\ref{fig:vcirc} 
for {\it all} DM models. This would exacerbate the tension of the 1~cm$^2$g$^{-1}$  SIDM model 
with the ultra-faint galaxies. 
It is however, possible for DM self-interactions to provide a novel explanation to the diversity we highlight here if the cross section is velocity dependent in such a way that it satisfies two conditions: (i) it is large enough to be above but near the threshold for gravothermal catastrophe at the typical internal velocities of MW satellites and (ii) it has a strong velocity dependence putting it well below this threshold at the orbital velocities of MW satellites within the MW. The former 
is required to have a fraction of SIDM subhalos collapse into cuspy density profiles, while the latter is required to avoid subhalo evaporation due to 
particles inside subhalos scattering with particles in the host halo, and it also minimizes the impact of self-interaction in the MW halo and beyond where constraints on the cross section are tight (e.g. \cite{Vogelsberger2012}). 

\begin{figure}
    \includegraphics[height=9.0cm,width=8.5cm, trim=2.05cm 0.5cm 1.25cm 0.0cm, clip=true]{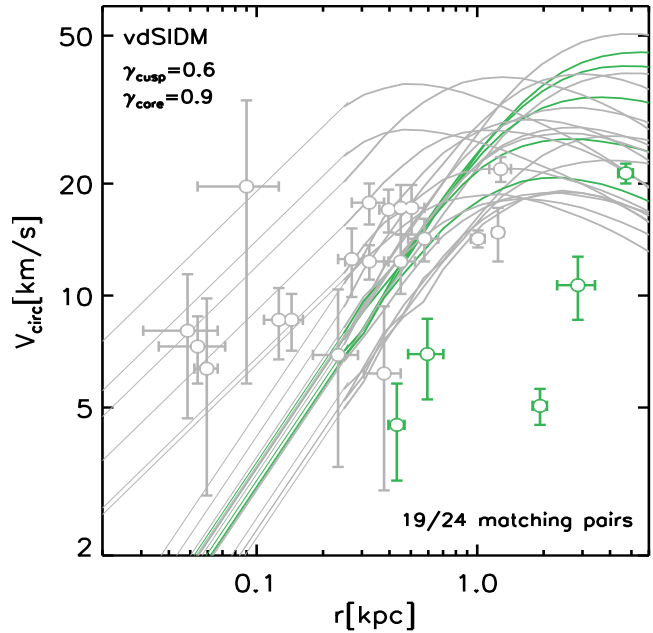}
\caption{As Fig.~\ref{fig:vcirc} but for the vdSIDM simulation (see Fig.~\ref{fig:models}). The velocity dependence of this SIDM model has cross sections near and above the onset of gravothermal
collapse for MW-like subhalos. This produces a bimodal subhalo distribution, with some of the systems developing a central cusp, $V_{\rm circ}\propto r^{0.6}$, while the others still retain a core, $V_{\rm circ}\propto r^{0.9}$.}
\label{fig:vcirc_vdsidm}
\end{figure}

The vdSIDM model we have explored (see Fig.~\ref{fig:models}) satisfies these requirements and is shown in Fig.~\ref{fig:vcirc_vdsidm}. We note that in this case we use the subhalo ranking according to the $V_{\rm max}$ values in the SIDM simulation, which has the same initial conditions. We do this because for those subhalos that have collapsed, the value of $V_{\rm max}$ changes substantially in the later epochs after accretion, and likely would not reflect the satellites' luminosity. The effect of the gravothermal collapse in this model is clear: it distinctly diversifies the subhalo population by producing a bimodal distribution, with low-mass subhalos being cuspy and offering a better match to the dense ultra-faint galaxies, while more massive subhalos remain cored and are better matched to the lower density satellites with large half-light radii. 

We can use the number of matching pairs as a way of ranking the five different (DM-only) models we have analysed in this work: 1) WDM (2.3 keV) with $21/24$, 2) vdSIDM and ETHOS-4 both with $19/24$, 3) SIDM ($1$~cm$^2$g$^{-1}$) with $16/24$ and 4) CDM with $15/24$. Another way of quantifying the difference between the models is the following. We compute the minimum chi-square of each model with {\it all} the satellites, i.e., for a given combination of 24 pairs (regardless of the number of matches), each being a subhalo circular velocity profile and a ($r_{1/2}$,$V_{1/2}$) observational point, we compute the chi-square of the combination\footnote{$$\chi^2=\sum_{i=1}^{24}\left(\frac{V_{\rm model}(r_{1/2}^i) - V_{1/2}^i)}{\sigma(V_{1/2}^i)}\right)^2$$ where $\sigma(V_{1/2}^i)$ is the observational error for a given $V_{1/2}^i$ data point.}, and explore a large number of combinations to find the minimum chi-square. To compare the models we compute the (reduced) chi-square difference $\Delta\chi^2_\nu$ ($\nu=24$) {\it relative to the WDM case}, which provides the best fit to the data, and find for each model: vdSIDM, $\Delta\chi^2_\nu\sim13.3~(3.6\sigma)$, ETHOS-4, $\Delta\chi^2_\nu\sim24.5~(5.0\sigma)$, SIDM ($1$~cm$^2$g$^{-1}$), $\Delta\chi^2_\nu\sim28.3~(5.3\sigma)$, and CDM, $\Delta\chi^2_\nu\sim42.5~(5.6\sigma)$. This comparison gives a similar ranking than the simple ordering based on the number of matching pairs, although in this case, the vdSIDM case fares better than the ETHOS-4 case. We emphasize that these rankings are merely indicative of the particular set of simulations we used and not a rigorous statistical test of the DM models given the limitations and circumstances of the simulations we have discussed in this section.

\section{Conclusions}\label{sec_conclusions}

The abundance and internal kinematics of MW satellites have been a challenge for the theory of structure formation. The discovery of ultra-faint galaxies coupled with the refinement of methods to estimate the DM mass within the half light radii is seemingly indicating that the MW satellites inhabit a subhalo population that has a strikingly diverse distribution of circular velocity profiles. 
In this work, we use DM-only simulations to highlight the potential implications of such diversity for the DM nature in models that deviate from the standard CDM model. Broadly we study classes of models having either a primordial power spectrum cutoff at galactic-scales (WDM) or strong DM self-interactions (SIDM) or a combination of both (as in the ETHOS framework \cite{Cyr2016,Vogelsberger2016}).  Our main conclusions are the following: 

\begin{itemize}[leftmargin=0.5cm]
\item[(i)] a primordial cutoff in the power spectrum suppresses the abundance of massive subhalos relative to CDM and naturally creates a more diverse subhalo population. A 2.3 keV WDM thermal relic model is quite consistent with the distribution of MW satellites but it is in tension with Ly-$\alpha$ forest constraints \cite{Irsic2017}.
\item[(ii)] DM self-interactions with $\sigma_T/m_\chi=1$~cm$^2$g$^{-1}$ are effective at reducing the DM density of MW subhalos at sub-kiloparsec scales but preserve the lack of diversity seen in CDM (without baryonic physics). Contrary to CDM, the predicted densities within $\sim100$~pc in SIDM are too low to be consistent with the ultra-faint galaxies. Although these galaxies have serious systematic uncertainties (see Section \ref{sec_uncertainties}), if their currently inferred high densities are verified, this poses a serious challenge for SIDM models with $\sigma/m_\chi\gtrsim1~$cm$^2$g$^{-1}$ 
at the velocity scales of MW satellites. 
\item[(iii)] we identify a novel way to explain the diversity of MW satellites and match the high densities of the ultra-faint galaxies within the SIDM model. It is based on the gravothermal collapse of SIDM halos and requires a velocity dependence in the SIDM cross section having a value above the collapse threshold at $v_{\rm rel}\sim60$~km/s (the velocity scale of massive MW satellites) and quickly dropping at higher velocities to avoid having an impact in the MW halo and larger systems where constraints on the cross section are tight. These conditions create a bimodal distribution of cored lower density subhalos more consistent with brighter satellites, and of cuspy dense subhalos consistent with the ultra-faint galaxies.
\item[(iv)] an allowed benchmark DM model within the ETHOS framework that has both a primordial power spectrum cutoff and DM self-interactions, and that has been shown to be a promising alternative to CDM \cite{Vogelsberger2016,Lovell2018}, is consistent with the ultra-faint galaxies and shows considerably more diversity than the CDM model (without baryonic physics).
\end{itemize}

We argue that including baryonic physics in the simulations would tend to increase the diversity of the subhalo population and likely lower the sub-kiloparsec DM densities (albeit see discussion at the end of section \ref{sec_discussion} for the SIDM case). 
Although quantifying this impact is a necessary task to constrain non-CDM models, we remark that based on this expectation, the challenge of SIDM in matching the kinematics of the ultra-faint galaxies would remain without a mechanism like the gravothermal collapse. In dwarfs more massive than the MW satellites ($V_{\rm max}>70$~km/s), how cuspy or cored SIDM halos are depends on the concentration of the baryonic component \cite{Elbert2018}. Indeed, the interplay between self-interactions and baryonic physics in this case has been invoked to address the observed large diversity of rotation curves for a constant cross section SIDM model with $\sigma_T/m_\chi\sim1$~cm$^2$g$^{-1}$ \cite{Kamada2017,Ren2018}. This interplay works because within the characteristic scale of these galaxies ($\gtrsim2$~kpc), the DM content is comparable to the baryonic content. However, at the scale of the ultra-faint galaxies ($\lesssim0.1$~kpc), the massive SIDM subhalos of the MW have enclosed masses $\sim3\times10^5$~M$_\odot$, which is $\sim100$ times more mass than the stellar mass of an ultra-faint galaxy. Hence the impact of baryonic physics in these systems should be much reduced, and it is difficult to imagine how a population of dense ultra-faint galaxies can be accommodated within SIDM with $\sigma_T/m_\chi\gtrsim1$~cm$^2$g$^{-1}$. 

\begin{acknowledgements}
JZ, MRL and JDB acknowledge support by a Grant of Excellence  from  the  Icelandic  Research  fund  (grant  number 173929$-$051). MRL is supported by a COFUND/Durham Junior Research Fellowship under EU grant 609412. The new simulation used for this work was performed on the joint MIT-Harvard computing cluster supported by MKI and FAS.
\end{acknowledgements}

\bibliography{diverse_MW_DM_nature}

\begin{thebibliography}{87}
\expandafter\ifx\csname natexlab\endcsname\relax\def\natexlab#1{#1}\fi
\expandafter\ifx\csname bibnamefont\endcsname\relax
  \def\bibnamefont#1{#1}\fi
\expandafter\ifx\csname bibfnamefont\endcsname\relax
  \def\bibfnamefont#1{#1}\fi
\expandafter\ifx\csname citenamefont\endcsname\relax
  \def\citenamefont#1{#1}\fi
\expandafter\ifx\csname url\endcsname\relax
  \def\url#1{\texttt{#1}}\fi
\expandafter\ifx\csname urlprefix\endcsname\relax\def\urlprefix{URL }\fi
\providecommand{\bibinfo}[2]{#2}
\providecommand{\eprint}[2][]{\url{#2}}

\bibitem[{\citenamefont{{Springel} et~al.}(2005)\citenamefont{{Springel},
  {White}, {Jenkins}, {Frenk}, {Yoshida}, {Gao}, {Navarro}, {Thacker},
  {Croton}, {Helly} et~al.}}]{Springel2005}
\bibinfo{author}{\bibfnamefont{V.}~\bibnamefont{{Springel}}},
  \bibinfo{author}{\bibfnamefont{S.~D.~M.} \bibnamefont{{White}}},
  \bibinfo{author}{\bibfnamefont{A.}~\bibnamefont{{Jenkins}}},
  \bibinfo{author}{\bibfnamefont{C.~S.} \bibnamefont{{Frenk}}},
  \bibinfo{author}{\bibfnamefont{N.}~\bibnamefont{{Yoshida}}},
  \bibinfo{author}{\bibfnamefont{L.}~\bibnamefont{{Gao}}},
  \bibinfo{author}{\bibfnamefont{J.}~\bibnamefont{{Navarro}}},
  \bibinfo{author}{\bibfnamefont{R.}~\bibnamefont{{Thacker}}},
  \bibinfo{author}{\bibfnamefont{D.}~\bibnamefont{{Croton}}},
  \bibinfo{author}{\bibfnamefont{J.}~\bibnamefont{{Helly}}},
  \bibnamefont{et~al.}, \bibinfo{journal}{\nat} \textbf{\bibinfo{volume}{435}},
  \bibinfo{pages}{629} (\bibinfo{year}{2005}), \eprint{astro-ph/0504097}.

\bibitem[{\citenamefont{{Dubois} et~al.}(2014)\citenamefont{{Dubois}, {Pichon},
  {Welker}, {Le Borgne}, {Devriendt}, {Laigle}, {Codis}, {Pogosyan}, {Arnouts},
  {Benabed} et~al.}}]{Dubois2014}
\bibinfo{author}{\bibfnamefont{Y.}~\bibnamefont{{Dubois}}},
  \bibinfo{author}{\bibfnamefont{C.}~\bibnamefont{{Pichon}}},
  \bibinfo{author}{\bibfnamefont{C.}~\bibnamefont{{Welker}}},
  \bibinfo{author}{\bibfnamefont{D.}~\bibnamefont{{Le Borgne}}},
  \bibinfo{author}{\bibfnamefont{J.}~\bibnamefont{{Devriendt}}},
  \bibinfo{author}{\bibfnamefont{C.}~\bibnamefont{{Laigle}}},
  \bibinfo{author}{\bibfnamefont{S.}~\bibnamefont{{Codis}}},
  \bibinfo{author}{\bibfnamefont{D.}~\bibnamefont{{Pogosyan}}},
  \bibinfo{author}{\bibfnamefont{S.}~\bibnamefont{{Arnouts}}},
  \bibinfo{author}{\bibfnamefont{K.}~\bibnamefont{{Benabed}}},
  \bibnamefont{et~al.}, \bibinfo{journal}{\mnras}
  \textbf{\bibinfo{volume}{444}}, \bibinfo{pages}{1453} (\bibinfo{year}{2014}),
  \eprint{1402.1165}.

\bibitem[{\citenamefont{{Vogelsberger}
  et~al.}(2014)\citenamefont{{Vogelsberger}, {Genel}, {Springel}, {Torrey},
  {Sijacki}, {Xu}, {Snyder}, {Nelson}, and {Hernquist}}}]{Vogelsberger2014}
\bibinfo{author}{\bibfnamefont{M.}~\bibnamefont{{Vogelsberger}}},
  \bibinfo{author}{\bibfnamefont{S.}~\bibnamefont{{Genel}}},
  \bibinfo{author}{\bibfnamefont{V.}~\bibnamefont{{Springel}}},
  \bibinfo{author}{\bibfnamefont{P.}~\bibnamefont{{Torrey}}},
  \bibinfo{author}{\bibfnamefont{D.}~\bibnamefont{{Sijacki}}},
  \bibinfo{author}{\bibfnamefont{D.}~\bibnamefont{{Xu}}},
  \bibinfo{author}{\bibfnamefont{G.}~\bibnamefont{{Snyder}}},
  \bibinfo{author}{\bibfnamefont{D.}~\bibnamefont{{Nelson}}}, \bibnamefont{and}
  \bibinfo{author}{\bibfnamefont{L.}~\bibnamefont{{Hernquist}}},
  \bibinfo{journal}{\mnras} \textbf{\bibinfo{volume}{444}},
  \bibinfo{pages}{1518} (\bibinfo{year}{2014}), \eprint{1405.2921}.

\bibitem[{\citenamefont{{Schaye} et~al.}(2015)\citenamefont{{Schaye}, {Crain},
  {Bower}, {Furlong}, {Schaller}, {Theuns}, {Dalla Vecchia}, {Frenk},
  {McCarthy}, {Helly} et~al.}}]{Schaye2015}
\bibinfo{author}{\bibfnamefont{J.}~\bibnamefont{{Schaye}}},
  \bibinfo{author}{\bibfnamefont{R.~A.} \bibnamefont{{Crain}}},
  \bibinfo{author}{\bibfnamefont{R.~G.} \bibnamefont{{Bower}}},
  \bibinfo{author}{\bibfnamefont{M.}~\bibnamefont{{Furlong}}},
  \bibinfo{author}{\bibfnamefont{M.}~\bibnamefont{{Schaller}}},
  \bibinfo{author}{\bibfnamefont{T.}~\bibnamefont{{Theuns}}},
  \bibinfo{author}{\bibfnamefont{C.}~\bibnamefont{{Dalla Vecchia}}},
  \bibinfo{author}{\bibfnamefont{C.~S.} \bibnamefont{{Frenk}}},
  \bibinfo{author}{\bibfnamefont{I.~G.} \bibnamefont{{McCarthy}}},
  \bibinfo{author}{\bibfnamefont{J.~C.} \bibnamefont{{Helly}}},
  \bibnamefont{et~al.}, \bibinfo{journal}{\mnras}
  \textbf{\bibinfo{volume}{446}}, \bibinfo{pages}{521} (\bibinfo{year}{2015}),
  \eprint{1407.7040}.

\bibitem[{\citenamefont{{Moore}}(1994)}]{Moore1994}
\bibinfo{author}{\bibfnamefont{B.}~\bibnamefont{{Moore}}},
  \bibinfo{journal}{\nat} \textbf{\bibinfo{volume}{370}}, \bibinfo{pages}{629}
  (\bibinfo{year}{1994}).

\bibitem[{\citenamefont{{Klypin} et~al.}(1999)\citenamefont{{Klypin},
  {Kravtsov}, {Valenzuela}, and {Prada}}}]{Klypin1999}
\bibinfo{author}{\bibfnamefont{A.}~\bibnamefont{{Klypin}}},
  \bibinfo{author}{\bibfnamefont{A.~V.} \bibnamefont{{Kravtsov}}},
  \bibinfo{author}{\bibfnamefont{O.}~\bibnamefont{{Valenzuela}}},
  \bibnamefont{and} \bibinfo{author}{\bibfnamefont{F.}~\bibnamefont{{Prada}}},
  \bibinfo{journal}{\apj} \textbf{\bibinfo{volume}{522}}, \bibinfo{pages}{82}
  (\bibinfo{year}{1999}), \eprint{astro-ph/9901240}.

\bibitem[{\citenamefont{{Moore} et~al.}(1999)\citenamefont{{Moore}, {Ghigna},
  {Governato}, {Lake}, {Quinn}, {Stadel}, and {Tozzi}}}]{Moore1999}
\bibinfo{author}{\bibfnamefont{B.}~\bibnamefont{{Moore}}},
  \bibinfo{author}{\bibfnamefont{S.}~\bibnamefont{{Ghigna}}},
  \bibinfo{author}{\bibfnamefont{F.}~\bibnamefont{{Governato}}},
  \bibinfo{author}{\bibfnamefont{G.}~\bibnamefont{{Lake}}},
  \bibinfo{author}{\bibfnamefont{T.}~\bibnamefont{{Quinn}}},
  \bibinfo{author}{\bibfnamefont{J.}~\bibnamefont{{Stadel}}}, \bibnamefont{and}
  \bibinfo{author}{\bibfnamefont{P.}~\bibnamefont{{Tozzi}}},
  \bibinfo{journal}{\apjl} \textbf{\bibinfo{volume}{524}}, \bibinfo{pages}{L19}
  (\bibinfo{year}{1999}), \eprint{astro-ph/9907411}.

\bibitem[{\citenamefont{{Zavala} et~al.}(2009)\citenamefont{{Zavala}, {Jing},
  {Faltenbacher}, {Yepes}, {Hoffman}, {Gottl{\"o}ber}, and
  {Catinella}}}]{Zavala2009}
\bibinfo{author}{\bibfnamefont{J.}~\bibnamefont{{Zavala}}},
  \bibinfo{author}{\bibfnamefont{Y.~P.} \bibnamefont{{Jing}}},
  \bibinfo{author}{\bibfnamefont{A.}~\bibnamefont{{Faltenbacher}}},
  \bibinfo{author}{\bibfnamefont{G.}~\bibnamefont{{Yepes}}},
  \bibinfo{author}{\bibfnamefont{Y.}~\bibnamefont{{Hoffman}}},
  \bibinfo{author}{\bibfnamefont{S.}~\bibnamefont{{Gottl{\"o}ber}}},
  \bibnamefont{and}
  \bibinfo{author}{\bibfnamefont{B.}~\bibnamefont{{Catinella}}},
  \bibinfo{journal}{\apj} \textbf{\bibinfo{volume}{700}}, \bibinfo{pages}{1779}
  (\bibinfo{year}{2009}), \eprint{0906.0585}.

\bibitem[{\citenamefont{{Walker} and {Pe{\~n}arrubia}}(2011)}]{Walker2011}
\bibinfo{author}{\bibfnamefont{M.~G.} \bibnamefont{{Walker}}} \bibnamefont{and}
  \bibinfo{author}{\bibfnamefont{J.}~\bibnamefont{{Pe{\~n}arrubia}}},
  \bibinfo{journal}{\apj} \textbf{\bibinfo{volume}{742}}, \bibinfo{eid}{20}
  (\bibinfo{year}{2011}), \eprint{1108.2404}.

\bibitem[{\citenamefont{{Boylan-Kolchin}
  et~al.}(2011)\citenamefont{{Boylan-Kolchin}, {Bullock}, and
  {Kaplinghat}}}]{Boylan-Kolchin2011}
\bibinfo{author}{\bibfnamefont{M.}~\bibnamefont{{Boylan-Kolchin}}},
  \bibinfo{author}{\bibfnamefont{J.~S.} \bibnamefont{{Bullock}}},
  \bibnamefont{and}
  \bibinfo{author}{\bibfnamefont{M.}~\bibnamefont{{Kaplinghat}}},
  \bibinfo{journal}{\mnras} \textbf{\bibinfo{volume}{415}},
  \bibinfo{pages}{L40} (\bibinfo{year}{2011}), \eprint{1103.0007}.

\bibitem[{\citenamefont{{Klypin} et~al.}(2014)\citenamefont{{Klypin},
  {Karachentsev}, {Makarov}, and {Nasonova}}}]{Klypin2014}
\bibinfo{author}{\bibfnamefont{A.}~\bibnamefont{{Klypin}}},
  \bibinfo{author}{\bibfnamefont{I.}~\bibnamefont{{Karachentsev}}},
  \bibinfo{author}{\bibfnamefont{D.}~\bibnamefont{{Makarov}}},
  \bibnamefont{and}
  \bibinfo{author}{\bibfnamefont{O.}~\bibnamefont{{Nasonova}}},
  \bibinfo{journal}{ArXiv e-prints}  (\bibinfo{year}{2014}),
  \eprint{1405.4523}.

\bibitem[{\citenamefont{{Papastergis} et~al.}(2015)\citenamefont{{Papastergis},
  {Giovanelli}, {Haynes}, and {Shankar}}}]{Papastergis2015}
\bibinfo{author}{\bibfnamefont{E.}~\bibnamefont{{Papastergis}}},
  \bibinfo{author}{\bibfnamefont{R.}~\bibnamefont{{Giovanelli}}},
  \bibinfo{author}{\bibfnamefont{M.~P.} \bibnamefont{{Haynes}}},
  \bibnamefont{and}
  \bibinfo{author}{\bibfnamefont{F.}~\bibnamefont{{Shankar}}},
  \bibinfo{journal}{\aap} \textbf{\bibinfo{volume}{574}}, \bibinfo{eid}{A113}
  (\bibinfo{year}{2015}), \eprint{1407.4665}.

\bibitem[{\citenamefont{{Oman} et~al.}(2015)\citenamefont{{Oman}, {Navarro},
  {Fattahi}, {Frenk}, {Sawala}, {White}, {Bower}, {Crain}, {Furlong},
  {Schaller} et~al.}}]{Oman2015}
\bibinfo{author}{\bibfnamefont{K.~A.} \bibnamefont{{Oman}}},
  \bibinfo{author}{\bibfnamefont{J.~F.} \bibnamefont{{Navarro}}},
  \bibinfo{author}{\bibfnamefont{A.}~\bibnamefont{{Fattahi}}},
  \bibinfo{author}{\bibfnamefont{C.~S.} \bibnamefont{{Frenk}}},
  \bibinfo{author}{\bibfnamefont{T.}~\bibnamefont{{Sawala}}},
  \bibinfo{author}{\bibfnamefont{S.~D.~M.} \bibnamefont{{White}}},
  \bibinfo{author}{\bibfnamefont{R.}~\bibnamefont{{Bower}}},
  \bibinfo{author}{\bibfnamefont{R.~A.} \bibnamefont{{Crain}}},
  \bibinfo{author}{\bibfnamefont{M.}~\bibnamefont{{Furlong}}},
  \bibinfo{author}{\bibfnamefont{M.}~\bibnamefont{{Schaller}}},
  \bibnamefont{et~al.}, \bibinfo{journal}{\mnras}
  \textbf{\bibinfo{volume}{452}}, \bibinfo{pages}{3650} (\bibinfo{year}{2015}),
  \eprint{1504.01437}.

\bibitem[{\citenamefont{{Pontzen} and {Governato}}(2012)}]{Pontzen2012}
\bibinfo{author}{\bibfnamefont{A.}~\bibnamefont{{Pontzen}}} \bibnamefont{and}
  \bibinfo{author}{\bibfnamefont{F.}~\bibnamefont{{Governato}}},
  \bibinfo{journal}{\mnras} \textbf{\bibinfo{volume}{421}},
  \bibinfo{pages}{3464} (\bibinfo{year}{2012}), \eprint{1106.0499}.

\bibitem[{\citenamefont{{Dutton} et~al.}(2016)\citenamefont{{Dutton},
  {Macci{\`o}}, {Frings}, {Wang}, {Stinson}, {Penzo}, and {Kang}}}]{Dutton2016}
\bibinfo{author}{\bibfnamefont{A.~A.} \bibnamefont{{Dutton}}},
  \bibinfo{author}{\bibfnamefont{A.~V.} \bibnamefont{{Macci{\`o}}}},
  \bibinfo{author}{\bibfnamefont{J.}~\bibnamefont{{Frings}}},
  \bibinfo{author}{\bibfnamefont{L.}~\bibnamefont{{Wang}}},
  \bibinfo{author}{\bibfnamefont{G.~S.} \bibnamefont{{Stinson}}},
  \bibinfo{author}{\bibfnamefont{C.}~\bibnamefont{{Penzo}}}, \bibnamefont{and}
  \bibinfo{author}{\bibfnamefont{X.}~\bibnamefont{{Kang}}},
  \bibinfo{journal}{\mnras} \textbf{\bibinfo{volume}{457}},
  \bibinfo{pages}{L74} (\bibinfo{year}{2016}), \eprint{1512.00453}.

\bibitem[{\citenamefont{{Sawala} et~al.}(2016)\citenamefont{{Sawala}, {Frenk},
  {Fattahi}, {Navarro}, {Bower}, {Crain}, {Dalla Vecchia}, {Furlong}, {Helly},
  {Jenkins} et~al.}}]{Sawala2016}
\bibinfo{author}{\bibfnamefont{T.}~\bibnamefont{{Sawala}}},
  \bibinfo{author}{\bibfnamefont{C.~S.} \bibnamefont{{Frenk}}},
  \bibinfo{author}{\bibfnamefont{A.}~\bibnamefont{{Fattahi}}},
  \bibinfo{author}{\bibfnamefont{J.~F.} \bibnamefont{{Navarro}}},
  \bibinfo{author}{\bibfnamefont{R.~G.} \bibnamefont{{Bower}}},
  \bibinfo{author}{\bibfnamefont{R.~A.} \bibnamefont{{Crain}}},
  \bibinfo{author}{\bibfnamefont{C.}~\bibnamefont{{Dalla Vecchia}}},
  \bibinfo{author}{\bibfnamefont{M.}~\bibnamefont{{Furlong}}},
  \bibinfo{author}{\bibfnamefont{J.~C.} \bibnamefont{{Helly}}},
  \bibinfo{author}{\bibfnamefont{A.}~\bibnamefont{{Jenkins}}},
  \bibnamefont{et~al.}, \bibinfo{journal}{\mnras}
  \textbf{\bibinfo{volume}{457}}, \bibinfo{pages}{1931} (\bibinfo{year}{2016}),
  \eprint{1511.01098}.

\bibitem[{\citenamefont{{Bullock} and {Boylan-Kolchin}}(2017)}]{Bullock2017}
\bibinfo{author}{\bibfnamefont{J.~S.} \bibnamefont{{Bullock}}}
  \bibnamefont{and}
  \bibinfo{author}{\bibfnamefont{M.}~\bibnamefont{{Boylan-Kolchin}}},
  \bibinfo{journal}{\araa} \textbf{\bibinfo{volume}{55}}, \bibinfo{pages}{343}
  (\bibinfo{year}{2017}), \eprint{1707.04256}.

\bibitem[{\citenamefont{{Col{\'{\i}}n}
  et~al.}(2000)\citenamefont{{Col{\'{\i}}n}, {Avila-Reese}, and
  {Valenzuela}}}]{Colin2000}
\bibinfo{author}{\bibfnamefont{P.}~\bibnamefont{{Col{\'{\i}}n}}},
  \bibinfo{author}{\bibfnamefont{V.}~\bibnamefont{{Avila-Reese}}},
  \bibnamefont{and}
  \bibinfo{author}{\bibfnamefont{O.}~\bibnamefont{{Valenzuela}}},
  \bibinfo{journal}{\apj} \textbf{\bibinfo{volume}{542}}, \bibinfo{pages}{622}
  (\bibinfo{year}{2000}), \eprint{astro-ph/0004115}.

\bibitem[{\citenamefont{{Bode} et~al.}(2001)\citenamefont{{Bode}, {Ostriker},
  and {Turok}}}]{Bode2001}
\bibinfo{author}{\bibfnamefont{P.}~\bibnamefont{{Bode}}},
  \bibinfo{author}{\bibfnamefont{J.~P.} \bibnamefont{{Ostriker}}},
  \bibnamefont{and} \bibinfo{author}{\bibfnamefont{N.}~\bibnamefont{{Turok}}},
  \bibinfo{journal}{\apj} \textbf{\bibinfo{volume}{556}}, \bibinfo{pages}{93}
  (\bibinfo{year}{2001}), \eprint{astro-ph/0010389}.

\bibitem[{\citenamefont{{B{\oe}hm} et~al.}(2002)\citenamefont{{B{\oe}hm},
  {Riazuelo}, {Hansen}, and {Schaeffer}}}]{Boehm2002}
\bibinfo{author}{\bibfnamefont{C.}~\bibnamefont{{B{\oe}hm}}},
  \bibinfo{author}{\bibfnamefont{A.}~\bibnamefont{{Riazuelo}}},
  \bibinfo{author}{\bibfnamefont{S.~H.} \bibnamefont{{Hansen}}},
  \bibnamefont{and}
  \bibinfo{author}{\bibfnamefont{R.}~\bibnamefont{{Schaeffer}}},
  \bibinfo{journal}{\prd} \textbf{\bibinfo{volume}{66}}, \bibinfo{eid}{083505}
  (\bibinfo{year}{2002}), \eprint{astro-ph/0112522}.

\bibitem[{\citenamefont{{Buckley} et~al.}(2014)\citenamefont{{Buckley},
  {Zavala}, {Cyr-Racine}, {Sigurdson}, and {Vogelsberger}}}]{Buckley2014}
\bibinfo{author}{\bibfnamefont{M.~R.} \bibnamefont{{Buckley}}},
  \bibinfo{author}{\bibfnamefont{J.}~\bibnamefont{{Zavala}}},
  \bibinfo{author}{\bibfnamefont{F.-Y.} \bibnamefont{{Cyr-Racine}}},
  \bibinfo{author}{\bibfnamefont{K.}~\bibnamefont{{Sigurdson}}},
  \bibnamefont{and}
  \bibinfo{author}{\bibfnamefont{M.}~\bibnamefont{{Vogelsberger}}},
  \bibinfo{journal}{\prd} \textbf{\bibinfo{volume}{90}}, \bibinfo{eid}{043524}
  (\bibinfo{year}{2014}), \eprint{1405.2075}.

\bibitem[{\citenamefont{{Spergel} and {Steinhardt}}(2000)}]{Spergel2000}
\bibinfo{author}{\bibfnamefont{D.~N.} \bibnamefont{{Spergel}}}
  \bibnamefont{and} \bibinfo{author}{\bibfnamefont{P.~J.}
  \bibnamefont{{Steinhardt}}}, \bibinfo{journal}{Physical Review Letters}
  \textbf{\bibinfo{volume}{84}}, \bibinfo{pages}{3760} (\bibinfo{year}{2000}),
  \eprint{astro-ph/9909386}.

\bibitem[{\citenamefont{{Lovell} et~al.}(2012)\citenamefont{{Lovell}, {Eke},
  {Frenk}, {Gao}, {Jenkins}, {Theuns}, {Wang}, {White}, {Boyarsky}, and
  {Ruchayskiy}}}]{Lovell2012}
\bibinfo{author}{\bibfnamefont{M.~R.} \bibnamefont{{Lovell}}},
  \bibinfo{author}{\bibfnamefont{V.}~\bibnamefont{{Eke}}},
  \bibinfo{author}{\bibfnamefont{C.~S.} \bibnamefont{{Frenk}}},
  \bibinfo{author}{\bibfnamefont{L.}~\bibnamefont{{Gao}}},
  \bibinfo{author}{\bibfnamefont{A.}~\bibnamefont{{Jenkins}}},
  \bibinfo{author}{\bibfnamefont{T.}~\bibnamefont{{Theuns}}},
  \bibinfo{author}{\bibfnamefont{J.}~\bibnamefont{{Wang}}},
  \bibinfo{author}{\bibfnamefont{S.~D.~M.} \bibnamefont{{White}}},
  \bibinfo{author}{\bibfnamefont{A.}~\bibnamefont{{Boyarsky}}},
  \bibnamefont{and}
  \bibinfo{author}{\bibfnamefont{O.}~\bibnamefont{{Ruchayskiy}}},
  \bibinfo{journal}{\mnras} \textbf{\bibinfo{volume}{420}},
  \bibinfo{pages}{2318} (\bibinfo{year}{2012}), \eprint{1104.2929}.

\bibitem[{\citenamefont{{Vogelsberger}
  et~al.}(2012)\citenamefont{{Vogelsberger}, {Zavala}, and
  {Loeb}}}]{Vogelsberger2012}
\bibinfo{author}{\bibfnamefont{M.}~\bibnamefont{{Vogelsberger}}},
  \bibinfo{author}{\bibfnamefont{J.}~\bibnamefont{{Zavala}}}, \bibnamefont{and}
  \bibinfo{author}{\bibfnamefont{A.}~\bibnamefont{{Loeb}}},
  \bibinfo{journal}{\mnras} \textbf{\bibinfo{volume}{423}},
  \bibinfo{pages}{3740} (\bibinfo{year}{2012}), \eprint{1201.5892}.

\bibitem[{\citenamefont{{Zavala} et~al.}(2013)\citenamefont{{Zavala},
  {Vogelsberger}, and {Walker}}}]{Zavala2013}
\bibinfo{author}{\bibfnamefont{J.}~\bibnamefont{{Zavala}}},
  \bibinfo{author}{\bibfnamefont{M.}~\bibnamefont{{Vogelsberger}}},
  \bibnamefont{and} \bibinfo{author}{\bibfnamefont{M.~G.}
  \bibnamefont{{Walker}}}, \bibinfo{journal}{\mnras}
  \textbf{\bibinfo{volume}{431}}, \bibinfo{pages}{L20} (\bibinfo{year}{2013}),
  \eprint{1211.6426}.

\bibitem[{\citenamefont{{Vogelsberger}
  et~al.}(2016)\citenamefont{{Vogelsberger}, {Zavala}, {Cyr-Racine},
  {Pfrommer}, {Bringmann}, and {Sigurdson}}}]{Vogelsberger2016}
\bibinfo{author}{\bibfnamefont{M.}~\bibnamefont{{Vogelsberger}}},
  \bibinfo{author}{\bibfnamefont{J.}~\bibnamefont{{Zavala}}},
  \bibinfo{author}{\bibfnamefont{F.-Y.} \bibnamefont{{Cyr-Racine}}},
  \bibinfo{author}{\bibfnamefont{C.}~\bibnamefont{{Pfrommer}}},
  \bibinfo{author}{\bibfnamefont{T.}~\bibnamefont{{Bringmann}}},
  \bibnamefont{and}
  \bibinfo{author}{\bibfnamefont{K.}~\bibnamefont{{Sigurdson}}},
  \bibinfo{journal}{\mnras} \textbf{\bibinfo{volume}{460}},
  \bibinfo{pages}{1399} (\bibinfo{year}{2016}), \eprint{1512.05349}.

\bibitem[{\citenamefont{{Schneider} et~al.}(2017)\citenamefont{{Schneider},
  {Trujillo-Gomez}, {Papastergis}, {Reed}, and {Lake}}}]{Schneider2017}
\bibinfo{author}{\bibfnamefont{A.}~\bibnamefont{{Schneider}}},
  \bibinfo{author}{\bibfnamefont{S.}~\bibnamefont{{Trujillo-Gomez}}},
  \bibinfo{author}{\bibfnamefont{E.}~\bibnamefont{{Papastergis}}},
  \bibinfo{author}{\bibfnamefont{D.~S.} \bibnamefont{{Reed}}},
  \bibnamefont{and} \bibinfo{author}{\bibfnamefont{G.}~\bibnamefont{{Lake}}},
  \bibinfo{journal}{\mnras} \textbf{\bibinfo{volume}{470}},
  \bibinfo{pages}{1542} (\bibinfo{year}{2017}), \eprint{1611.09362}.

\bibitem[{\citenamefont{{Kamada} et~al.}(2017)\citenamefont{{Kamada},
  {Kaplinghat}, {Pace}, and {Yu}}}]{Kamada2017}
\bibinfo{author}{\bibfnamefont{A.}~\bibnamefont{{Kamada}}},
  \bibinfo{author}{\bibfnamefont{M.}~\bibnamefont{{Kaplinghat}}},
  \bibinfo{author}{\bibfnamefont{A.~B.} \bibnamefont{{Pace}}},
  \bibnamefont{and} \bibinfo{author}{\bibfnamefont{H.-B.} \bibnamefont{{Yu}}},
  \bibinfo{journal}{Physical Review Letters} \textbf{\bibinfo{volume}{119}},
  \bibinfo{eid}{111102} (\bibinfo{year}{2017}), \eprint{1611.02716}.

\bibitem[{\citenamefont{{Burger} and {Zavala}}(2019)}]{Burger2019}
\bibinfo{author}{\bibfnamefont{J.~D.} \bibnamefont{{Burger}}} \bibnamefont{and}
  \bibinfo{author}{\bibfnamefont{J.}~\bibnamefont{{Zavala}}},
  \bibinfo{journal}{\mnras} \textbf{\bibinfo{volume}{485}},
  \bibinfo{pages}{1008} (\bibinfo{year}{2019}), \eprint{1810.10024}.

\bibitem[{\citenamefont{{Boylan-Kolchin}
  et~al.}(2012{\natexlab{a}})\citenamefont{{Boylan-Kolchin}, {Bullock}, and
  {Kaplinghat}}}]{Boylan-Kolchin2012}
\bibinfo{author}{\bibfnamefont{M.}~\bibnamefont{{Boylan-Kolchin}}},
  \bibinfo{author}{\bibfnamefont{J.~S.} \bibnamefont{{Bullock}}},
  \bibnamefont{and}
  \bibinfo{author}{\bibfnamefont{M.}~\bibnamefont{{Kaplinghat}}},
  \bibinfo{journal}{\mnras} \textbf{\bibinfo{volume}{422}},
  \bibinfo{pages}{1203} (\bibinfo{year}{2012}{\natexlab{a}}),
  \eprint{1111.2048}.

\bibitem[{\citenamefont{{Fattahi} et~al.}(2018)\citenamefont{{Fattahi},
  {Navarro}, {Frenk}, {Oman}, {Sawala}, and {Schaller}}}]{Fattahi2018}
\bibinfo{author}{\bibfnamefont{A.}~\bibnamefont{{Fattahi}}},
  \bibinfo{author}{\bibfnamefont{J.~F.} \bibnamefont{{Navarro}}},
  \bibinfo{author}{\bibfnamefont{C.~S.} \bibnamefont{{Frenk}}},
  \bibinfo{author}{\bibfnamefont{K.~A.} \bibnamefont{{Oman}}},
  \bibinfo{author}{\bibfnamefont{T.}~\bibnamefont{{Sawala}}}, \bibnamefont{and}
  \bibinfo{author}{\bibfnamefont{M.}~\bibnamefont{{Schaller}}},
  \bibinfo{journal}{\mnras} \textbf{\bibinfo{volume}{476}},
  \bibinfo{pages}{3816} (\bibinfo{year}{2018}), \eprint{1707.03898}.

\bibitem[{\citenamefont{{Errani} et~al.}(2018)\citenamefont{{Errani},
  {Pe{\~n}arrubia}, and {Walker}}}]{Errani2018}
\bibinfo{author}{\bibfnamefont{R.}~\bibnamefont{{Errani}}},
  \bibinfo{author}{\bibfnamefont{J.}~\bibnamefont{{Pe{\~n}arrubia}}},
  \bibnamefont{and} \bibinfo{author}{\bibfnamefont{M.~G.}
  \bibnamefont{{Walker}}}, \bibinfo{journal}{ArXiv e-prints}
  (\bibinfo{year}{2018}), \eprint{1805.00484}.

\bibitem[{\citenamefont{{Springel} et~al.}(2008)\citenamefont{{Springel},
  {Wang}, {Vogelsberger}, {Ludlow}, {Jenkins}, {Helmi}, {Navarro}, {Frenk}, and
  {White}}}]{Springel2008}
\bibinfo{author}{\bibfnamefont{V.}~\bibnamefont{{Springel}}},
  \bibinfo{author}{\bibfnamefont{J.}~\bibnamefont{{Wang}}},
  \bibinfo{author}{\bibfnamefont{M.}~\bibnamefont{{Vogelsberger}}},
  \bibinfo{author}{\bibfnamefont{A.}~\bibnamefont{{Ludlow}}},
  \bibinfo{author}{\bibfnamefont{A.}~\bibnamefont{{Jenkins}}},
  \bibinfo{author}{\bibfnamefont{A.}~\bibnamefont{{Helmi}}},
  \bibinfo{author}{\bibfnamefont{J.~F.} \bibnamefont{{Navarro}}},
  \bibinfo{author}{\bibfnamefont{C.~S.} \bibnamefont{{Frenk}}},
  \bibnamefont{and} \bibinfo{author}{\bibfnamefont{S.~D.~M.}
  \bibnamefont{{White}}}, \bibinfo{journal}{\mnras}
  \textbf{\bibinfo{volume}{391}}, \bibinfo{pages}{1685} (\bibinfo{year}{2008}),
  \eprint{0809.0898}.

\bibitem[{\citenamefont{{Lovell} et~al.}(2014)\citenamefont{{Lovell}, {Frenk},
  {Eke}, {Jenkins}, {Gao}, and {Theuns}}}]{Lovell2014}
\bibinfo{author}{\bibfnamefont{M.~R.} \bibnamefont{{Lovell}}},
  \bibinfo{author}{\bibfnamefont{C.~S.} \bibnamefont{{Frenk}}},
  \bibinfo{author}{\bibfnamefont{V.~R.} \bibnamefont{{Eke}}},
  \bibinfo{author}{\bibfnamefont{A.}~\bibnamefont{{Jenkins}}},
  \bibinfo{author}{\bibfnamefont{L.}~\bibnamefont{{Gao}}}, \bibnamefont{and}
  \bibinfo{author}{\bibfnamefont{T.}~\bibnamefont{{Theuns}}},
  \bibinfo{journal}{\mnras} \textbf{\bibinfo{volume}{439}},
  \bibinfo{pages}{300} (\bibinfo{year}{2014}), \eprint{1308.1399}.

\bibitem[{\citenamefont{{Spergel} et~al.}(2003)\citenamefont{{Spergel},
  {Verde}, {Peiris}, {Komatsu}, {Nolta}, {Bennett}, {Halpern}, {Hinshaw},
  {Jarosik}, {Kogut} et~al.}}]{Spergel2003}
\bibinfo{author}{\bibfnamefont{D.~N.} \bibnamefont{{Spergel}}},
  \bibinfo{author}{\bibfnamefont{L.}~\bibnamefont{{Verde}}},
  \bibinfo{author}{\bibfnamefont{H.~V.} \bibnamefont{{Peiris}}},
  \bibinfo{author}{\bibfnamefont{E.}~\bibnamefont{{Komatsu}}},
  \bibinfo{author}{\bibfnamefont{M.~R.} \bibnamefont{{Nolta}}},
  \bibinfo{author}{\bibfnamefont{C.~L.} \bibnamefont{{Bennett}}},
  \bibinfo{author}{\bibfnamefont{M.}~\bibnamefont{{Halpern}}},
  \bibinfo{author}{\bibfnamefont{G.}~\bibnamefont{{Hinshaw}}},
  \bibinfo{author}{\bibfnamefont{N.}~\bibnamefont{{Jarosik}}},
  \bibinfo{author}{\bibfnamefont{A.}~\bibnamefont{{Kogut}}},
  \bibnamefont{et~al.}, \bibinfo{journal}{\apjs}
  \textbf{\bibinfo{volume}{148}}, \bibinfo{pages}{175} (\bibinfo{year}{2003}),
  \eprint{astro-ph/0302209}.

\bibitem[{\citenamefont{{Komatsu} et~al.}(2011)\citenamefont{{Komatsu},
  {Smith}, {Dunkley}, {Bennett}, {Gold}, {Hinshaw}, {Jarosik}, {Larson},
  {Nolta}, {Page} et~al.}}]{Komatsu2011}
\bibinfo{author}{\bibfnamefont{E.}~\bibnamefont{{Komatsu}}},
  \bibinfo{author}{\bibfnamefont{K.~M.} \bibnamefont{{Smith}}},
  \bibinfo{author}{\bibfnamefont{J.}~\bibnamefont{{Dunkley}}},
  \bibinfo{author}{\bibfnamefont{C.~L.} \bibnamefont{{Bennett}}},
  \bibinfo{author}{\bibfnamefont{B.}~\bibnamefont{{Gold}}},
  \bibinfo{author}{\bibfnamefont{G.}~\bibnamefont{{Hinshaw}}},
  \bibinfo{author}{\bibfnamefont{N.}~\bibnamefont{{Jarosik}}},
  \bibinfo{author}{\bibfnamefont{D.}~\bibnamefont{{Larson}}},
  \bibinfo{author}{\bibfnamefont{M.~R.} \bibnamefont{{Nolta}}},
  \bibinfo{author}{\bibfnamefont{L.}~\bibnamefont{{Page}}},
  \bibnamefont{et~al.}, \bibinfo{journal}{\apjs}
  \textbf{\bibinfo{volume}{192}}, \bibinfo{eid}{18} (\bibinfo{year}{2011}),
  \eprint{1001.4538}.

\bibitem[{\citenamefont{{Planck Collaboration}
  et~al.}(2016)\citenamefont{{Planck Collaboration}, {Ade}, {Aghanim},
  {Arnaud}, {Ashdown}, {Aumont}, {Baccigalupi}, {Banday}, {Barreiro},
  {Bartlett} et~al.}}]{Planck2016}
\bibinfo{author}{\bibnamefont{{Planck Collaboration}}},
  \bibinfo{author}{\bibfnamefont{P.~A.~R.} \bibnamefont{{Ade}}},
  \bibinfo{author}{\bibfnamefont{N.}~\bibnamefont{{Aghanim}}},
  \bibinfo{author}{\bibfnamefont{M.}~\bibnamefont{{Arnaud}}},
  \bibinfo{author}{\bibfnamefont{M.}~\bibnamefont{{Ashdown}}},
  \bibinfo{author}{\bibfnamefont{J.}~\bibnamefont{{Aumont}}},
  \bibinfo{author}{\bibfnamefont{C.}~\bibnamefont{{Baccigalupi}}},
  \bibinfo{author}{\bibfnamefont{A.~J.} \bibnamefont{{Banday}}},
  \bibinfo{author}{\bibfnamefont{R.~B.} \bibnamefont{{Barreiro}}},
  \bibinfo{author}{\bibfnamefont{J.~G.} \bibnamefont{{Bartlett}}},
  \bibnamefont{et~al.}, \bibinfo{journal}{\aap} \textbf{\bibinfo{volume}{594}},
  \bibinfo{eid}{A13} (\bibinfo{year}{2016}), \eprint{1502.01589}.

\bibitem[{\citenamefont{{Lovell}
  et~al.}(2018{\natexlab{a}})\citenamefont{{Lovell}, {Zavala}, and
  {Vogelsberger}}}]{Lovell2019}
\bibinfo{author}{\bibfnamefont{M.~R.} \bibnamefont{{Lovell}}},
  \bibinfo{author}{\bibfnamefont{J.}~\bibnamefont{{Zavala}}}, \bibnamefont{and}
  \bibinfo{author}{\bibfnamefont{M.}~\bibnamefont{{Vogelsberger}}},
  \bibinfo{journal}{arXiv e-prints}  (\bibinfo{year}{2018}{\natexlab{a}}),
  \eprint{1812.04627}.

\bibitem[{\citenamefont{{Viel} et~al.}(2013)\citenamefont{{Viel}, {Becker},
  {Bolton}, and {Haehnelt}}}]{Viel2013}
\bibinfo{author}{\bibfnamefont{M.}~\bibnamefont{{Viel}}},
  \bibinfo{author}{\bibfnamefont{G.~D.} \bibnamefont{{Becker}}},
  \bibinfo{author}{\bibfnamefont{J.~S.} \bibnamefont{{Bolton}}},
  \bibnamefont{and} \bibinfo{author}{\bibfnamefont{M.~G.}
  \bibnamefont{{Haehnelt}}}, \bibinfo{journal}{\prd}
  \textbf{\bibinfo{volume}{88}}, \bibinfo{eid}{043502} (\bibinfo{year}{2013}),
  \eprint{1306.2314}.

\bibitem[{\citenamefont{{Peter} et~al.}(2013)\citenamefont{{Peter}, {Rocha},
  {Bullock}, and {Kaplinghat}}}]{Peter2013}
\bibinfo{author}{\bibfnamefont{A.~H.~G.} \bibnamefont{{Peter}}},
  \bibinfo{author}{\bibfnamefont{M.}~\bibnamefont{{Rocha}}},
  \bibinfo{author}{\bibfnamefont{J.~S.} \bibnamefont{{Bullock}}},
  \bibnamefont{and}
  \bibinfo{author}{\bibfnamefont{M.}~\bibnamefont{{Kaplinghat}}},
  \bibinfo{journal}{\mnras} \textbf{\bibinfo{volume}{430}},
  \bibinfo{pages}{105} (\bibinfo{year}{2013}), \eprint{1208.3026}.

\bibitem[{\citenamefont{{Cyr-Racine} et~al.}(2016)\citenamefont{{Cyr-Racine},
  {Sigurdson}, {Zavala}, {Bringmann}, {Vogelsberger}, and
  {Pfrommer}}}]{Cyr2016}
\bibinfo{author}{\bibfnamefont{F.-Y.} \bibnamefont{{Cyr-Racine}}},
  \bibinfo{author}{\bibfnamefont{K.}~\bibnamefont{{Sigurdson}}},
  \bibinfo{author}{\bibfnamefont{J.}~\bibnamefont{{Zavala}}},
  \bibinfo{author}{\bibfnamefont{T.}~\bibnamefont{{Bringmann}}},
  \bibinfo{author}{\bibfnamefont{M.}~\bibnamefont{{Vogelsberger}}},
  \bibnamefont{and}
  \bibinfo{author}{\bibfnamefont{C.}~\bibnamefont{{Pfrommer}}},
  \bibinfo{journal}{\prd} \textbf{\bibinfo{volume}{93}}, \bibinfo{eid}{123527}
  (\bibinfo{year}{2016}), \eprint{1512.05344}.

\bibitem[{\citenamefont{{Ir{\v s}i{\v c}} et~al.}(2017)\citenamefont{{Ir{\v
  s}i{\v c}}, {Viel}, {Haehnelt}, {Bolton}, {Cristiani}, {Becker}, {D'Odorico},
  {Cupani}, {Kim}, {Berg} et~al.}}]{Irsic2017}
\bibinfo{author}{\bibfnamefont{V.}~\bibnamefont{{Ir{\v s}i{\v c}}}},
  \bibinfo{author}{\bibfnamefont{M.}~\bibnamefont{{Viel}}},
  \bibinfo{author}{\bibfnamefont{M.~G.} \bibnamefont{{Haehnelt}}},
  \bibinfo{author}{\bibfnamefont{J.~S.} \bibnamefont{{Bolton}}},
  \bibinfo{author}{\bibfnamefont{S.}~\bibnamefont{{Cristiani}}},
  \bibinfo{author}{\bibfnamefont{G.~D.} \bibnamefont{{Becker}}},
  \bibinfo{author}{\bibfnamefont{V.}~\bibnamefont{{D'Odorico}}},
  \bibinfo{author}{\bibfnamefont{G.}~\bibnamefont{{Cupani}}},
  \bibinfo{author}{\bibfnamefont{T.-S.} \bibnamefont{{Kim}}},
  \bibinfo{author}{\bibfnamefont{T.~A.~M.} \bibnamefont{{Berg}}},
  \bibnamefont{et~al.}, \bibinfo{journal}{\prd} \textbf{\bibinfo{volume}{96}},
  \bibinfo{eid}{023522} (\bibinfo{year}{2017}), \eprint{1702.01764}.

\bibitem[{\citenamefont{{Garzilli} et~al.}(2017)\citenamefont{{Garzilli},
  {Boyarsky}, and {Ruchayskiy}}}]{Garzilli2017}
\bibinfo{author}{\bibfnamefont{A.}~\bibnamefont{{Garzilli}}},
  \bibinfo{author}{\bibfnamefont{A.}~\bibnamefont{{Boyarsky}}},
  \bibnamefont{and}
  \bibinfo{author}{\bibfnamefont{O.}~\bibnamefont{{Ruchayskiy}}},
  \bibinfo{journal}{Physics Letters B} \textbf{\bibinfo{volume}{773}},
  \bibinfo{pages}{258} (\bibinfo{year}{2017}), \eprint{1510.07006}.

\bibitem[{\citenamefont{{Walther} et~al.}(2018)\citenamefont{{Walther},
  {O{\~n}orbe}, {Hennawi}, and {Luki{\'c}}}}]{Walther2018}
\bibinfo{author}{\bibfnamefont{M.}~\bibnamefont{{Walther}}},
  \bibinfo{author}{\bibfnamefont{J.}~\bibnamefont{{O{\~n}orbe}}},
  \bibinfo{author}{\bibfnamefont{J.~F.} \bibnamefont{{Hennawi}}},
  \bibnamefont{and}
  \bibinfo{author}{\bibfnamefont{Z.}~\bibnamefont{{Luki{\'c}}}},
  \bibinfo{journal}{ArXiv e-prints}  (\bibinfo{year}{2018}),
  \eprint{1808.04367}.

\bibitem[{\citenamefont{{Bose} et~al.}(2018)\citenamefont{{Bose},
  {Vogelsberger}, {Zavala}, {Pfrommer}, {Cyr-Racine}, {Bohr}, and
  {Bringmann}}}]{Bose2018}
\bibinfo{author}{\bibfnamefont{S.}~\bibnamefont{{Bose}}},
  \bibinfo{author}{\bibfnamefont{M.}~\bibnamefont{{Vogelsberger}}},
  \bibinfo{author}{\bibfnamefont{J.}~\bibnamefont{{Zavala}}},
  \bibinfo{author}{\bibfnamefont{C.}~\bibnamefont{{Pfrommer}}},
  \bibinfo{author}{\bibfnamefont{F.-Y.} \bibnamefont{{Cyr-Racine}}},
  \bibinfo{author}{\bibfnamefont{S.}~\bibnamefont{{Bohr}}}, \bibnamefont{and}
  \bibinfo{author}{\bibfnamefont{T.}~\bibnamefont{{Bringmann}}},
  \bibinfo{journal}{arXiv e-prints}  (\bibinfo{year}{2018}),
  \eprint{1811.10630}.

\bibitem[{\citenamefont{{Tulin} and {Yu}}(2018)}]{Tulin2018}
\bibinfo{author}{\bibfnamefont{S.}~\bibnamefont{{Tulin}}} \bibnamefont{and}
  \bibinfo{author}{\bibfnamefont{H.-B.} \bibnamefont{{Yu}}},
  \bibinfo{journal}{\physrep} \textbf{\bibinfo{volume}{730}},
  \bibinfo{pages}{1} (\bibinfo{year}{2018}), \eprint{1705.02358}.

\bibitem[{\citenamefont{{Read} et~al.}(2018)\citenamefont{{Read}, {Walker}, and
  {Steger}}}]{Read2018}
\bibinfo{author}{\bibfnamefont{J.~I.} \bibnamefont{{Read}}},
  \bibinfo{author}{\bibfnamefont{M.~G.} \bibnamefont{{Walker}}},
  \bibnamefont{and} \bibinfo{author}{\bibfnamefont{P.}~\bibnamefont{{Steger}}},
  \bibinfo{journal}{ArXiv e-prints}  (\bibinfo{year}{2018}),
  \eprint{1805.06934}.

\bibitem[{\citenamefont{{Valli} and {Yu}}(2018)}]{Valli2018}
\bibinfo{author}{\bibfnamefont{M.}~\bibnamefont{{Valli}}} \bibnamefont{and}
  \bibinfo{author}{\bibfnamefont{H.-B.} \bibnamefont{{Yu}}},
  \bibinfo{journal}{Nature Astronomy} \textbf{\bibinfo{volume}{2}},
  \bibinfo{pages}{907} (\bibinfo{year}{2018}), \eprint{1711.03502}.

\bibitem[{\citenamefont{{Robles} et~al.}(2017)\citenamefont{{Robles},
  {Bullock}, {Elbert}, {Fitts}, {Gonz{\'a}lez-Samaniego}, {Boylan-Kolchin},
  {Hopkins}, {Faucher-Gigu{\`e}re}, {Kere{\v s}}, and {Hayward}}}]{Robles2017}
\bibinfo{author}{\bibfnamefont{V.~H.} \bibnamefont{{Robles}}},
  \bibinfo{author}{\bibfnamefont{J.~S.} \bibnamefont{{Bullock}}},
  \bibinfo{author}{\bibfnamefont{O.~D.} \bibnamefont{{Elbert}}},
  \bibinfo{author}{\bibfnamefont{A.}~\bibnamefont{{Fitts}}},
  \bibinfo{author}{\bibfnamefont{A.}~\bibnamefont{{Gonz{\'a}lez-Samaniego}}},
  \bibinfo{author}{\bibfnamefont{M.}~\bibnamefont{{Boylan-Kolchin}}},
  \bibinfo{author}{\bibfnamefont{P.~F.} \bibnamefont{{Hopkins}}},
  \bibinfo{author}{\bibfnamefont{C.-A.} \bibnamefont{{Faucher-Gigu{\`e}re}}},
  \bibinfo{author}{\bibfnamefont{D.}~\bibnamefont{{Kere{\v s}}}},
  \bibnamefont{and} \bibinfo{author}{\bibfnamefont{C.~C.}
  \bibnamefont{{Hayward}}}, \bibinfo{journal}{\mnras}
  \textbf{\bibinfo{volume}{472}}, \bibinfo{pages}{2945} (\bibinfo{year}{2017}),
  \eprint{1706.07514}.

\bibitem[{\citenamefont{{Feng} et~al.}(2010)\citenamefont{{Feng}, {Kaplinghat},
  and {Yu}}}]{Feng2010}
\bibinfo{author}{\bibfnamefont{J.~L.} \bibnamefont{{Feng}}},
  \bibinfo{author}{\bibfnamefont{M.}~\bibnamefont{{Kaplinghat}}},
  \bibnamefont{and} \bibinfo{author}{\bibfnamefont{H.-B.} \bibnamefont{{Yu}}},
  \bibinfo{journal}{Physical Review Letters} \textbf{\bibinfo{volume}{104}},
  \bibinfo{eid}{151301} (\bibinfo{year}{2010}), \eprint{0911.0422}.

\bibitem[{\citenamefont{{Loeb} and {Weiner}}(2011)}]{Loeb2011}
\bibinfo{author}{\bibfnamefont{A.}~\bibnamefont{{Loeb}}} \bibnamefont{and}
  \bibinfo{author}{\bibfnamefont{N.}~\bibnamefont{{Weiner}}},
  \bibinfo{journal}{Physical Review Letters} \textbf{\bibinfo{volume}{106}},
  \bibinfo{eid}{171302} (\bibinfo{year}{2011}), \eprint{1011.6374}.

\bibitem[{\citenamefont{{Lynden-Bell} and {Wood}}(1968)}]{LyndenBell1968}
\bibinfo{author}{\bibfnamefont{D.}~\bibnamefont{{Lynden-Bell}}}
  \bibnamefont{and} \bibinfo{author}{\bibfnamefont{R.}~\bibnamefont{{Wood}}},
  \bibinfo{journal}{\mnras} \textbf{\bibinfo{volume}{138}},
  \bibinfo{pages}{495} (\bibinfo{year}{1968}).

\bibitem[{\citenamefont{{Balberg} et~al.}(2002)\citenamefont{{Balberg},
  {Shapiro}, and {Inagaki}}}]{Balberg2002}
\bibinfo{author}{\bibfnamefont{S.}~\bibnamefont{{Balberg}}},
  \bibinfo{author}{\bibfnamefont{S.~L.} \bibnamefont{{Shapiro}}},
  \bibnamefont{and}
  \bibinfo{author}{\bibfnamefont{S.}~\bibnamefont{{Inagaki}}},
  \bibinfo{journal}{\apj} \textbf{\bibinfo{volume}{568}}, \bibinfo{pages}{475}
  (\bibinfo{year}{2002}), \eprint{astro-ph/0110561}.

\bibitem[{\citenamefont{{Col{\'\i}n} et~al.}(2002)\citenamefont{{Col{\'\i}n},
  {Avila-Reese}, {Valenzuela}, and {Firmani}}}]{Colin2002}
\bibinfo{author}{\bibfnamefont{P.}~\bibnamefont{{Col{\'\i}n}}},
  \bibinfo{author}{\bibfnamefont{V.}~\bibnamefont{{Avila-Reese}}},
  \bibinfo{author}{\bibfnamefont{O.}~\bibnamefont{{Valenzuela}}},
  \bibnamefont{and}
  \bibinfo{author}{\bibfnamefont{C.}~\bibnamefont{{Firmani}}},
  \bibinfo{journal}{\apj} \textbf{\bibinfo{volume}{581}}, \bibinfo{pages}{777}
  (\bibinfo{year}{2002}), \eprint{astro-ph/0205322}.

\bibitem[{\citenamefont{{Koda} and {Shapiro}}(2011)}]{Koda2011}
\bibinfo{author}{\bibfnamefont{J.}~\bibnamefont{{Koda}}} \bibnamefont{and}
  \bibinfo{author}{\bibfnamefont{P.~R.} \bibnamefont{{Shapiro}}},
  \bibinfo{journal}{\mnras} \textbf{\bibinfo{volume}{415}},
  \bibinfo{pages}{1125} (\bibinfo{year}{2011}), \eprint{1101.3097}.

\bibitem[{\citenamefont{{Pollack} et~al.}(2015)\citenamefont{{Pollack},
  {Spergel}, and {Steinhardt}}}]{Pollack2015}
\bibinfo{author}{\bibfnamefont{J.}~\bibnamefont{{Pollack}}},
  \bibinfo{author}{\bibfnamefont{D.~N.} \bibnamefont{{Spergel}}},
  \bibnamefont{and} \bibinfo{author}{\bibfnamefont{P.~J.}
  \bibnamefont{{Steinhardt}}}, \bibinfo{journal}{\apj}
  \textbf{\bibinfo{volume}{804}}, \bibinfo{eid}{131} (\bibinfo{year}{2015}),
  \eprint{1501.00017}.

\bibitem[{\citenamefont{{Nishikawa} et~al.}(2019)\citenamefont{{Nishikawa},
  {Boddy}, and {Kaplinghat}}}]{Nishikawa2019}
\bibinfo{author}{\bibfnamefont{H.}~\bibnamefont{{Nishikawa}}},
  \bibinfo{author}{\bibfnamefont{K.~K.} \bibnamefont{{Boddy}}},
  \bibnamefont{and}
  \bibinfo{author}{\bibfnamefont{M.}~\bibnamefont{{Kaplinghat}}},
  \bibinfo{journal}{arXiv e-prints}  (\bibinfo{year}{2019}),
  \eprint{1901.00499}.

\bibitem[{\citenamefont{{Torrealba} et~al.}(2018)\citenamefont{{Torrealba},
  {Belokurov}, {Koposov}, {Li}, {Walker}, {Sanders}, {Geringer-Sameth},
  {Zucker}, {Kuehn}, {Evans} et~al.}}]{Torrealba2018}
\bibinfo{author}{\bibfnamefont{G.}~\bibnamefont{{Torrealba}}},
  \bibinfo{author}{\bibfnamefont{V.}~\bibnamefont{{Belokurov}}},
  \bibinfo{author}{\bibfnamefont{S.~E.} \bibnamefont{{Koposov}}},
  \bibinfo{author}{\bibfnamefont{T.~S.} \bibnamefont{{Li}}},
  \bibinfo{author}{\bibfnamefont{M.~G.} \bibnamefont{{Walker}}},
  \bibinfo{author}{\bibfnamefont{J.~L.} \bibnamefont{{Sanders}}},
  \bibinfo{author}{\bibfnamefont{A.}~\bibnamefont{{Geringer-Sameth}}},
  \bibinfo{author}{\bibfnamefont{D.~B.} \bibnamefont{{Zucker}}},
  \bibinfo{author}{\bibfnamefont{K.}~\bibnamefont{{Kuehn}}},
  \bibinfo{author}{\bibfnamefont{N.~W.} \bibnamefont{{Evans}}},
  \bibnamefont{et~al.}, \bibinfo{journal}{arXiv e-prints}
  (\bibinfo{year}{2018}), \eprint{1811.04082}.

\bibitem[{\citenamefont{{Elbert} et~al.}(2015)\citenamefont{{Elbert},
  {Bullock}, {Garrison-Kimmel}, {Rocha}, {O{\~n}orbe}, and
  {Peter}}}]{Elbert2015}
\bibinfo{author}{\bibfnamefont{O.~D.} \bibnamefont{{Elbert}}},
  \bibinfo{author}{\bibfnamefont{J.~S.} \bibnamefont{{Bullock}}},
  \bibinfo{author}{\bibfnamefont{S.}~\bibnamefont{{Garrison-Kimmel}}},
  \bibinfo{author}{\bibfnamefont{M.}~\bibnamefont{{Rocha}}},
  \bibinfo{author}{\bibfnamefont{J.}~\bibnamefont{{O{\~n}orbe}}},
  \bibnamefont{and} \bibinfo{author}{\bibfnamefont{A.~H.~G.}
  \bibnamefont{{Peter}}}, \bibinfo{journal}{\mnras}
  \textbf{\bibinfo{volume}{453}}, \bibinfo{pages}{29} (\bibinfo{year}{2015}),
  \eprint{1412.1477}.

\bibitem[{\citenamefont{{Polisensky} and {Ricotti}}(2014)}]{Polisensky2014}
\bibinfo{author}{\bibfnamefont{E.}~\bibnamefont{{Polisensky}}}
  \bibnamefont{and}
  \bibinfo{author}{\bibfnamefont{M.}~\bibnamefont{{Ricotti}}},
  \bibinfo{journal}{\mnras} \textbf{\bibinfo{volume}{437}},
  \bibinfo{pages}{2922} (\bibinfo{year}{2014}), \eprint{1310.0430}.

\bibitem[{\citenamefont{{Wang} et~al.}(2015)\citenamefont{{Wang}, {Han},
  {Cooper}, {Cole}, {Frenk}, and {Lowing}}}]{Wang2015}
\bibinfo{author}{\bibfnamefont{W.}~\bibnamefont{{Wang}}},
  \bibinfo{author}{\bibfnamefont{J.}~\bibnamefont{{Han}}},
  \bibinfo{author}{\bibfnamefont{A.~P.} \bibnamefont{{Cooper}}},
  \bibinfo{author}{\bibfnamefont{S.}~\bibnamefont{{Cole}}},
  \bibinfo{author}{\bibfnamefont{C.}~\bibnamefont{{Frenk}}}, \bibnamefont{and}
  \bibinfo{author}{\bibfnamefont{B.}~\bibnamefont{{Lowing}}},
  \bibinfo{journal}{\mnras} \textbf{\bibinfo{volume}{453}},
  \bibinfo{pages}{377} (\bibinfo{year}{2015}), \eprint{1502.03477}.

\bibitem[{\citenamefont{{Patel} et~al.}(2018)\citenamefont{{Patel}, {Besla},
  {Mandel}, and {Sohn}}}]{Patel2018}
\bibinfo{author}{\bibfnamefont{E.}~\bibnamefont{{Patel}}},
  \bibinfo{author}{\bibfnamefont{G.}~\bibnamefont{{Besla}}},
  \bibinfo{author}{\bibfnamefont{K.}~\bibnamefont{{Mandel}}}, \bibnamefont{and}
  \bibinfo{author}{\bibfnamefont{S.~T.} \bibnamefont{{Sohn}}},
  \bibinfo{journal}{\apj} \textbf{\bibinfo{volume}{857}}, \bibinfo{eid}{78}
  (\bibinfo{year}{2018}), \eprint{1803.01878}.

\bibitem[{\citenamefont{{Boylan-Kolchin}
  et~al.}(2010)\citenamefont{{Boylan-Kolchin}, {Springel}, {White}, and
  {Jenkins}}}]{BK2010}
\bibinfo{author}{\bibfnamefont{M.}~\bibnamefont{{Boylan-Kolchin}}},
  \bibinfo{author}{\bibfnamefont{V.}~\bibnamefont{{Springel}}},
  \bibinfo{author}{\bibfnamefont{S.~D.~M.} \bibnamefont{{White}}},
  \bibnamefont{and}
  \bibinfo{author}{\bibfnamefont{A.}~\bibnamefont{{Jenkins}}},
  \bibinfo{journal}{\mnras} \textbf{\bibinfo{volume}{406}},
  \bibinfo{pages}{896} (\bibinfo{year}{2010}), \eprint{0911.4484}.

\bibitem[{\citenamefont{{Cautun} et~al.}(2014)\citenamefont{{Cautun},
  {Hellwing}, {van de Weygaert}, {Frenk}, {Jones}, and {Sawala}}}]{Cautun2014}
\bibinfo{author}{\bibfnamefont{M.}~\bibnamefont{{Cautun}}},
  \bibinfo{author}{\bibfnamefont{W.~A.} \bibnamefont{{Hellwing}}},
  \bibinfo{author}{\bibfnamefont{R.}~\bibnamefont{{van de Weygaert}}},
  \bibinfo{author}{\bibfnamefont{C.~S.} \bibnamefont{{Frenk}}},
  \bibinfo{author}{\bibfnamefont{B.~J.~T.} \bibnamefont{{Jones}}},
  \bibnamefont{and} \bibinfo{author}{\bibfnamefont{T.}~\bibnamefont{{Sawala}}},
  \bibinfo{journal}{\mnras} \textbf{\bibinfo{volume}{445}},
  \bibinfo{pages}{1820} (\bibinfo{year}{2014}), \eprint{1405.7700}.

\bibitem[{\citenamefont{{Guo} et~al.}(2015)\citenamefont{{Guo}, {Cooper},
  {Frenk}, {Helly}, and {Hellwing}}}]{Guo2015}
\bibinfo{author}{\bibfnamefont{Q.}~\bibnamefont{{Guo}}},
  \bibinfo{author}{\bibfnamefont{A.~P.} \bibnamefont{{Cooper}}},
  \bibinfo{author}{\bibfnamefont{C.}~\bibnamefont{{Frenk}}},
  \bibinfo{author}{\bibfnamefont{J.}~\bibnamefont{{Helly}}}, \bibnamefont{and}
  \bibinfo{author}{\bibfnamefont{W.~A.} \bibnamefont{{Hellwing}}},
  \bibinfo{journal}{\mnras} \textbf{\bibinfo{volume}{454}},
  \bibinfo{pages}{550} (\bibinfo{year}{2015}), \eprint{1503.08508}.

\bibitem[{\citenamefont{{Boylan-Kolchin}
  et~al.}(2012{\natexlab{b}})\citenamefont{{Boylan-Kolchin}, {Bullock}, and
  {Kaplinghat}}}]{BK2012}
\bibinfo{author}{\bibfnamefont{M.}~\bibnamefont{{Boylan-Kolchin}}},
  \bibinfo{author}{\bibfnamefont{J.~S.} \bibnamefont{{Bullock}}},
  \bibnamefont{and}
  \bibinfo{author}{\bibfnamefont{M.}~\bibnamefont{{Kaplinghat}}},
  \bibinfo{journal}{\mnras} \textbf{\bibinfo{volume}{422}},
  \bibinfo{pages}{1203} (\bibinfo{year}{2012}{\natexlab{b}}),
  \eprint{1111.2048}.

\bibitem[{\citenamefont{{Barnes} and {White}}(1984)}]{Barnes1984}
\bibinfo{author}{\bibfnamefont{J.}~\bibnamefont{{Barnes}}} \bibnamefont{and}
  \bibinfo{author}{\bibfnamefont{S.~D.~M.} \bibnamefont{{White}}},
  \bibinfo{journal}{\mnras} \textbf{\bibinfo{volume}{211}},
  \bibinfo{pages}{753} (\bibinfo{year}{1984}).

\bibitem[{\citenamefont{{Blumenthal} et~al.}(1986)\citenamefont{{Blumenthal},
  {Faber}, {Flores}, and {Primack}}}]{Blu1986}
\bibinfo{author}{\bibfnamefont{G.~R.} \bibnamefont{{Blumenthal}}},
  \bibinfo{author}{\bibfnamefont{S.~M.} \bibnamefont{{Faber}}},
  \bibinfo{author}{\bibfnamefont{R.}~\bibnamefont{{Flores}}}, \bibnamefont{and}
  \bibinfo{author}{\bibfnamefont{J.~R.} \bibnamefont{{Primack}}},
  \bibinfo{journal}{\apj} \textbf{\bibinfo{volume}{301}}, \bibinfo{pages}{27}
  (\bibinfo{year}{1986}).

\bibitem[{\citenamefont{{Di Cintio} et~al.}(2014)\citenamefont{{Di Cintio},
  {Brook}, {Macci{\`o}}, {Stinson}, {Knebe}, {Dutton}, and
  {Wadsley}}}]{diCintio2014}
\bibinfo{author}{\bibfnamefont{A.}~\bibnamefont{{Di Cintio}}},
  \bibinfo{author}{\bibfnamefont{C.~B.} \bibnamefont{{Brook}}},
  \bibinfo{author}{\bibfnamefont{A.~V.} \bibnamefont{{Macci{\`o}}}},
  \bibinfo{author}{\bibfnamefont{G.~S.} \bibnamefont{{Stinson}}},
  \bibinfo{author}{\bibfnamefont{A.}~\bibnamefont{{Knebe}}},
  \bibinfo{author}{\bibfnamefont{A.~A.} \bibnamefont{{Dutton}}},
  \bibnamefont{and}
  \bibinfo{author}{\bibfnamefont{J.}~\bibnamefont{{Wadsley}}},
  \bibinfo{journal}{\mnras} \textbf{\bibinfo{volume}{437}},
  \bibinfo{pages}{415} (\bibinfo{year}{2014}), \eprint{1306.0898}.

\bibitem[{\citenamefont{{Zolotov} et~al.}(2012)\citenamefont{{Zolotov},
  {Brooks}, {Willman}, {Governato}, {Pontzen}, {Christensen}, {Dekel}, {Quinn},
  {Shen}, and {Wadsley}}}]{Zolotov2012}
\bibinfo{author}{\bibfnamefont{A.}~\bibnamefont{{Zolotov}}},
  \bibinfo{author}{\bibfnamefont{A.~M.} \bibnamefont{{Brooks}}},
  \bibinfo{author}{\bibfnamefont{B.}~\bibnamefont{{Willman}}},
  \bibinfo{author}{\bibfnamefont{F.}~\bibnamefont{{Governato}}},
  \bibinfo{author}{\bibfnamefont{A.}~\bibnamefont{{Pontzen}}},
  \bibinfo{author}{\bibfnamefont{C.}~\bibnamefont{{Christensen}}},
  \bibinfo{author}{\bibfnamefont{A.}~\bibnamefont{{Dekel}}},
  \bibinfo{author}{\bibfnamefont{T.}~\bibnamefont{{Quinn}}},
  \bibinfo{author}{\bibfnamefont{S.}~\bibnamefont{{Shen}}}, \bibnamefont{and}
  \bibinfo{author}{\bibfnamefont{J.}~\bibnamefont{{Wadsley}}},
  \bibinfo{journal}{\apj} \textbf{\bibinfo{volume}{761}}, \bibinfo{eid}{71}
  (\bibinfo{year}{2012}), \eprint{1207.0007}.

\bibitem[{\citenamefont{{Garrison-Kimmel}
  et~al.}(2018)\citenamefont{{Garrison-Kimmel}, {Hopkins}, {Wetzel}, {Bullock},
  {Boylan-Kolchin}, {Keres}, {Faucher-Giguere}, {El-Badry}, {Lamberts},
  {Quataert} et~al.}}]{GK2018}
\bibinfo{author}{\bibfnamefont{S.}~\bibnamefont{{Garrison-Kimmel}}},
  \bibinfo{author}{\bibfnamefont{P.~F.} \bibnamefont{{Hopkins}}},
  \bibinfo{author}{\bibfnamefont{A.}~\bibnamefont{{Wetzel}}},
  \bibinfo{author}{\bibfnamefont{J.~S.} \bibnamefont{{Bullock}}},
  \bibinfo{author}{\bibfnamefont{M.}~\bibnamefont{{Boylan-Kolchin}}},
  \bibinfo{author}{\bibfnamefont{D.}~\bibnamefont{{Keres}}},
  \bibinfo{author}{\bibfnamefont{C.-A.} \bibnamefont{{Faucher-Giguere}}},
  \bibinfo{author}{\bibfnamefont{K.}~\bibnamefont{{El-Badry}}},
  \bibinfo{author}{\bibfnamefont{A.}~\bibnamefont{{Lamberts}}},
  \bibinfo{author}{\bibfnamefont{E.}~\bibnamefont{{Quataert}}},
  \bibnamefont{et~al.}, \bibinfo{journal}{ArXiv e-prints}
  (\bibinfo{year}{2018}), \eprint{1806.04143}.

\bibitem[{\citenamefont{{Garrison-Kimmel}
  et~al.}(2017)\citenamefont{{Garrison-Kimmel}, {Wetzel}, {Bullock}, {Hopkins},
  {Boylan-Kolchin}, {Faucher-Gigu{\`e}re}, {Kere{\v s}}, {Quataert},
  {Sanderson}, {Graus} et~al.}}]{GK2017}
\bibinfo{author}{\bibfnamefont{S.}~\bibnamefont{{Garrison-Kimmel}}},
  \bibinfo{author}{\bibfnamefont{A.}~\bibnamefont{{Wetzel}}},
  \bibinfo{author}{\bibfnamefont{J.~S.} \bibnamefont{{Bullock}}},
  \bibinfo{author}{\bibfnamefont{P.~F.} \bibnamefont{{Hopkins}}},
  \bibinfo{author}{\bibfnamefont{M.}~\bibnamefont{{Boylan-Kolchin}}},
  \bibinfo{author}{\bibfnamefont{C.-A.} \bibnamefont{{Faucher-Gigu{\`e}re}}},
  \bibinfo{author}{\bibfnamefont{D.}~\bibnamefont{{Kere{\v s}}}},
  \bibinfo{author}{\bibfnamefont{E.}~\bibnamefont{{Quataert}}},
  \bibinfo{author}{\bibfnamefont{R.~E.} \bibnamefont{{Sanderson}}},
  \bibinfo{author}{\bibfnamefont{A.~S.} \bibnamefont{{Graus}}},
  \bibnamefont{et~al.}, \bibinfo{journal}{\mnras}
  \textbf{\bibinfo{volume}{471}}, \bibinfo{pages}{1709} (\bibinfo{year}{2017}),
  \eprint{1701.03792}.

\bibitem[{\citenamefont{{Lovell} et~al.}(2017)\citenamefont{{Lovell},
  {Gonzalez-Perez}, {Bose}, {Boyarsky}, {Cole}, {Frenk}, and
  {Ruchayskiy}}}]{Lovell2017}
\bibinfo{author}{\bibfnamefont{M.~R.} \bibnamefont{{Lovell}}},
  \bibinfo{author}{\bibfnamefont{V.}~\bibnamefont{{Gonzalez-Perez}}},
  \bibinfo{author}{\bibfnamefont{S.}~\bibnamefont{{Bose}}},
  \bibinfo{author}{\bibfnamefont{A.}~\bibnamefont{{Boyarsky}}},
  \bibinfo{author}{\bibfnamefont{S.}~\bibnamefont{{Cole}}},
  \bibinfo{author}{\bibfnamefont{C.~S.} \bibnamefont{{Frenk}}},
  \bibnamefont{and}
  \bibinfo{author}{\bibfnamefont{O.}~\bibnamefont{{Ruchayskiy}}},
  \bibinfo{journal}{\mnras} \textbf{\bibinfo{volume}{468}},
  \bibinfo{pages}{2836} (\bibinfo{year}{2017}), \eprint{1611.00005}.

\bibitem[{\citenamefont{{Robles} et~al.}(2019)\citenamefont{{Robles}, {Kelley},
  {Bullock}, and {Kaplinghat}}}]{Robles2019}
\bibinfo{author}{\bibfnamefont{V.~H.} \bibnamefont{{Robles}}},
  \bibinfo{author}{\bibfnamefont{T.}~\bibnamefont{{Kelley}}},
  \bibinfo{author}{\bibfnamefont{J.~S.} \bibnamefont{{Bullock}}},
  \bibnamefont{and}
  \bibinfo{author}{\bibfnamefont{M.}~\bibnamefont{{Kaplinghat}}},
  \bibinfo{journal}{arXiv e-prints}  (\bibinfo{year}{2019}),
  \eprint{1903.01469}.

\bibitem[{\citenamefont{{Kahlhoefer} et~al.}(2019)\citenamefont{{Kahlhoefer},
  {Kaplinghat}, {Slatyer}, and {Wu}}}]{Kahlhoefer2019}
\bibinfo{author}{\bibfnamefont{F.}~\bibnamefont{{Kahlhoefer}}},
  \bibinfo{author}{\bibfnamefont{M.}~\bibnamefont{{Kaplinghat}}},
  \bibinfo{author}{\bibfnamefont{T.~R.} \bibnamefont{{Slatyer}}},
  \bibnamefont{and} \bibinfo{author}{\bibfnamefont{C.-L.} \bibnamefont{{Wu}}},
  \bibinfo{journal}{arXiv e-prints} \bibinfo{eid}{arXiv:1904.10539}
  (\bibinfo{year}{2019}), \eprint{1904.10539}.

\bibitem[{\citenamefont{{Sameie} et~al.}(2019)\citenamefont{{Sameie}, {Yu},
  {Sales}, {Vogelsberger}, and {Zavala}}}]{Sameie2019}
\bibinfo{author}{\bibfnamefont{O.}~\bibnamefont{{Sameie}}},
  \bibinfo{author}{\bibfnamefont{H.-B.} \bibnamefont{{Yu}}},
  \bibinfo{author}{\bibfnamefont{L.~V.} \bibnamefont{{Sales}}},
  \bibinfo{author}{\bibfnamefont{M.}~\bibnamefont{{Vogelsberger}}},
  \bibnamefont{and} \bibinfo{author}{\bibfnamefont{J.}~\bibnamefont{{Zavala}}},
  \bibinfo{journal}{arXiv e-prints} \bibinfo{eid}{arXiv:1904.07872}
  (\bibinfo{year}{2019}), \eprint{1904.07872}.

\bibitem[{\citenamefont{{Simon} et~al.}(2011)\citenamefont{{Simon}, {Geha},
  {Minor}, {Martinez}, {Kirby}, {Bullock}, {Kaplinghat}, {Strigari}, {Willman},
  {Choi} et~al.}}]{Simon2011}
\bibinfo{author}{\bibfnamefont{J.~D.} \bibnamefont{{Simon}}},
  \bibinfo{author}{\bibfnamefont{M.}~\bibnamefont{{Geha}}},
  \bibinfo{author}{\bibfnamefont{Q.~E.} \bibnamefont{{Minor}}},
  \bibinfo{author}{\bibfnamefont{G.~D.} \bibnamefont{{Martinez}}},
  \bibinfo{author}{\bibfnamefont{E.~N.} \bibnamefont{{Kirby}}},
  \bibinfo{author}{\bibfnamefont{J.~S.} \bibnamefont{{Bullock}}},
  \bibinfo{author}{\bibfnamefont{M.}~\bibnamefont{{Kaplinghat}}},
  \bibinfo{author}{\bibfnamefont{L.~E.} \bibnamefont{{Strigari}}},
  \bibinfo{author}{\bibfnamefont{B.}~\bibnamefont{{Willman}}},
  \bibinfo{author}{\bibfnamefont{P.~I.} \bibnamefont{{Choi}}},
  \bibnamefont{et~al.}, \bibinfo{journal}{\apj} \textbf{\bibinfo{volume}{733}},
  \bibinfo{eid}{46} (\bibinfo{year}{2011}), \eprint{1007.4198}.

\bibitem[{\citenamefont{{Bonnivard} et~al.}(2016)\citenamefont{{Bonnivard},
  {Maurin}, and {Walker}}}]{Bonnivard2016}
\bibinfo{author}{\bibfnamefont{V.}~\bibnamefont{{Bonnivard}}},
  \bibinfo{author}{\bibfnamefont{D.}~\bibnamefont{{Maurin}}}, \bibnamefont{and}
  \bibinfo{author}{\bibfnamefont{M.~G.} \bibnamefont{{Walker}}},
  \bibinfo{journal}{\mnras} \textbf{\bibinfo{volume}{462}},
  \bibinfo{pages}{223} (\bibinfo{year}{2016}), \eprint{1506.08209}.

\bibitem[{\citenamefont{{Martin} et~al.}(2007)\citenamefont{{Martin}, {Ibata},
  {Chapman}, {Irwin}, and {Lewis}}}]{Martin2007}
\bibinfo{author}{\bibfnamefont{N.~F.} \bibnamefont{{Martin}}},
  \bibinfo{author}{\bibfnamefont{R.~A.} \bibnamefont{{Ibata}}},
  \bibinfo{author}{\bibfnamefont{S.~C.} \bibnamefont{{Chapman}}},
  \bibinfo{author}{\bibfnamefont{M.}~\bibnamefont{{Irwin}}}, \bibnamefont{and}
  \bibinfo{author}{\bibfnamefont{G.~F.} \bibnamefont{{Lewis}}},
  \bibinfo{journal}{\mnras} \textbf{\bibinfo{volume}{380}},
  \bibinfo{pages}{281} (\bibinfo{year}{2007}), \eprint{0705.4622}.

\bibitem[{\citenamefont{{Willman} et~al.}(2011)\citenamefont{{Willman}, {Geha},
  {Strader}, {Strigari}, {Simon}, {Kirby}, {Ho}, and {Warres}}}]{Willman2011}
\bibinfo{author}{\bibfnamefont{B.}~\bibnamefont{{Willman}}},
  \bibinfo{author}{\bibfnamefont{M.}~\bibnamefont{{Geha}}},
  \bibinfo{author}{\bibfnamefont{J.}~\bibnamefont{{Strader}}},
  \bibinfo{author}{\bibfnamefont{L.~E.} \bibnamefont{{Strigari}}},
  \bibinfo{author}{\bibfnamefont{J.~D.} \bibnamefont{{Simon}}},
  \bibinfo{author}{\bibfnamefont{E.}~\bibnamefont{{Kirby}}},
  \bibinfo{author}{\bibfnamefont{N.}~\bibnamefont{{Ho}}}, \bibnamefont{and}
  \bibinfo{author}{\bibfnamefont{A.}~\bibnamefont{{Warres}}},
  \bibinfo{journal}{\aj} \textbf{\bibinfo{volume}{142}}, \bibinfo{eid}{128}
  (\bibinfo{year}{2011}), \eprint{1007.3499}.

\bibitem[{\citenamefont{{Belokurov} et~al.}(2009)\citenamefont{{Belokurov},
  {Walker}, {Evans}, {Gilmore}, {Irwin}, {Mateo}, {Mayer}, {Olszewski},
  {Bechtold}, and {Pickering}}}]{Belokurov2009}
\bibinfo{author}{\bibfnamefont{V.}~\bibnamefont{{Belokurov}}},
  \bibinfo{author}{\bibfnamefont{M.~G.} \bibnamefont{{Walker}}},
  \bibinfo{author}{\bibfnamefont{N.~W.} \bibnamefont{{Evans}}},
  \bibinfo{author}{\bibfnamefont{G.}~\bibnamefont{{Gilmore}}},
  \bibinfo{author}{\bibfnamefont{M.~J.} \bibnamefont{{Irwin}}},
  \bibinfo{author}{\bibfnamefont{M.}~\bibnamefont{{Mateo}}},
  \bibinfo{author}{\bibfnamefont{L.}~\bibnamefont{{Mayer}}},
  \bibinfo{author}{\bibfnamefont{E.}~\bibnamefont{{Olszewski}}},
  \bibinfo{author}{\bibfnamefont{J.}~\bibnamefont{{Bechtold}}},
  \bibnamefont{and}
  \bibinfo{author}{\bibfnamefont{T.}~\bibnamefont{{Pickering}}},
  \bibinfo{journal}{\mnras} \textbf{\bibinfo{volume}{397}},
  \bibinfo{pages}{1748} (\bibinfo{year}{2009}), \eprint{0903.0818}.

\bibitem[{\citenamefont{{Koch} et~al.}(2009)\citenamefont{{Koch}, {Wilkinson},
  {Kleyna}, {Irwin}, {Zucker}, {Belokurov}, {Gilmore}, {Fellhauer}, and
  {Evans}}}]{Koch2009}
\bibinfo{author}{\bibfnamefont{A.}~\bibnamefont{{Koch}}},
  \bibinfo{author}{\bibfnamefont{M.~I.} \bibnamefont{{Wilkinson}}},
  \bibinfo{author}{\bibfnamefont{J.~T.} \bibnamefont{{Kleyna}}},
  \bibinfo{author}{\bibfnamefont{M.}~\bibnamefont{{Irwin}}},
  \bibinfo{author}{\bibfnamefont{D.~B.} \bibnamefont{{Zucker}}},
  \bibinfo{author}{\bibfnamefont{V.}~\bibnamefont{{Belokurov}}},
  \bibinfo{author}{\bibfnamefont{G.~F.} \bibnamefont{{Gilmore}}},
  \bibinfo{author}{\bibfnamefont{M.}~\bibnamefont{{Fellhauer}}},
  \bibnamefont{and} \bibinfo{author}{\bibfnamefont{N.~W.}
  \bibnamefont{{Evans}}}, \bibinfo{journal}{\apj}
  \textbf{\bibinfo{volume}{690}}, \bibinfo{pages}{453} (\bibinfo{year}{2009}),
  \eprint{0809.0700}.

\bibitem[{\citenamefont{{Kirby} et~al.}(2013)\citenamefont{{Kirby},
  {Boylan-Kolchin}, {Cohen}, {Geha}, {Bullock}, and {Kaplinghat}}}]{Kirby2013}
\bibinfo{author}{\bibfnamefont{E.~N.} \bibnamefont{{Kirby}}},
  \bibinfo{author}{\bibfnamefont{M.}~\bibnamefont{{Boylan-Kolchin}}},
  \bibinfo{author}{\bibfnamefont{J.~G.} \bibnamefont{{Cohen}}},
  \bibinfo{author}{\bibfnamefont{M.}~\bibnamefont{{Geha}}},
  \bibinfo{author}{\bibfnamefont{J.~S.} \bibnamefont{{Bullock}}},
  \bibnamefont{and}
  \bibinfo{author}{\bibfnamefont{M.}~\bibnamefont{{Kaplinghat}}},
  \bibinfo{journal}{\apj} \textbf{\bibinfo{volume}{770}}, \bibinfo{eid}{16}
  (\bibinfo{year}{2013}), \eprint{1304.6080}.

\bibitem[{\citenamefont{{Ji} et~al.}(2016)\citenamefont{{Ji}, {Frebel},
  {Simon}, and {Geha}}}]{Ji2016}
\bibinfo{author}{\bibfnamefont{A.~P.} \bibnamefont{{Ji}}},
  \bibinfo{author}{\bibfnamefont{A.}~\bibnamefont{{Frebel}}},
  \bibinfo{author}{\bibfnamefont{J.~D.} \bibnamefont{{Simon}}},
  \bibnamefont{and} \bibinfo{author}{\bibfnamefont{M.}~\bibnamefont{{Geha}}},
  \bibinfo{journal}{\apj} \textbf{\bibinfo{volume}{817}}, \bibinfo{eid}{41}
  (\bibinfo{year}{2016}), \eprint{1510.07632}.

\bibitem[{\citenamefont{{Lovell}
  et~al.}(2018{\natexlab{b}})\citenamefont{{Lovell}, {Zavala}, {Vogelsberger},
  {Shen}, {Cyr-Racine}, {Pfrommer}, {Sigurdson}, {Boylan-Kolchin}, and
  {Pillepich}}}]{Lovell2018}
\bibinfo{author}{\bibfnamefont{M.~R.} \bibnamefont{{Lovell}}},
  \bibinfo{author}{\bibfnamefont{J.}~\bibnamefont{{Zavala}}},
  \bibinfo{author}{\bibfnamefont{M.}~\bibnamefont{{Vogelsberger}}},
  \bibinfo{author}{\bibfnamefont{X.}~\bibnamefont{{Shen}}},
  \bibinfo{author}{\bibfnamefont{F.-Y.} \bibnamefont{{Cyr-Racine}}},
  \bibinfo{author}{\bibfnamefont{C.}~\bibnamefont{{Pfrommer}}},
  \bibinfo{author}{\bibfnamefont{K.}~\bibnamefont{{Sigurdson}}},
  \bibinfo{author}{\bibfnamefont{M.}~\bibnamefont{{Boylan-Kolchin}}},
  \bibnamefont{and}
  \bibinfo{author}{\bibfnamefont{A.}~\bibnamefont{{Pillepich}}},
  \bibinfo{journal}{\mnras} \textbf{\bibinfo{volume}{477}},
  \bibinfo{pages}{2886} (\bibinfo{year}{2018}{\natexlab{b}}),
  \eprint{1711.10497}.

\bibitem[{\citenamefont{{Elbert} et~al.}(2018)\citenamefont{{Elbert},
  {Bullock}, {Kaplinghat}, {Garrison-Kimmel}, {Graus}, and
  {Rocha}}}]{Elbert2018}
\bibinfo{author}{\bibfnamefont{O.~D.} \bibnamefont{{Elbert}}},
  \bibinfo{author}{\bibfnamefont{J.~S.} \bibnamefont{{Bullock}}},
  \bibinfo{author}{\bibfnamefont{M.}~\bibnamefont{{Kaplinghat}}},
  \bibinfo{author}{\bibfnamefont{S.}~\bibnamefont{{Garrison-Kimmel}}},
  \bibinfo{author}{\bibfnamefont{A.~S.} \bibnamefont{{Graus}}},
  \bibnamefont{and} \bibinfo{author}{\bibfnamefont{M.}~\bibnamefont{{Rocha}}},
  \bibinfo{journal}{\apj} \textbf{\bibinfo{volume}{853}}, \bibinfo{eid}{109}
  (\bibinfo{year}{2018}), \eprint{1609.08626}.

\bibitem[{\citenamefont{{Ren} et~al.}(2018)\citenamefont{{Ren}, {Kwa},
  {Kaplinghat}, and {Yu}}}]{Ren2018}
\bibinfo{author}{\bibfnamefont{T.}~\bibnamefont{{Ren}}},
  \bibinfo{author}{\bibfnamefont{A.}~\bibnamefont{{Kwa}}},
  \bibinfo{author}{\bibfnamefont{M.}~\bibnamefont{{Kaplinghat}}},
  \bibnamefont{and} \bibinfo{author}{\bibfnamefont{H.-B.} \bibnamefont{{Yu}}},
  \bibinfo{journal}{ArXiv e-prints}  (\bibinfo{year}{2018}),
  \eprint{1808.05695}.

\end{thebibliography}

\end{document}